\documentclass[final,5p,times,twocolumn]{elsarticle}
\usepackage[utf8]{inputenc}
\usepackage[T1]{fontenc}
\usepackage{graphicx}
\usepackage{grffile}
\usepackage{longtable}
\usepackage{wrapfig}
\usepackage{rotating}
\usepackage[normalem]{ulem}
\usepackage{amsmath}
\usepackage{textcomp}
\usepackage{amssymb}
\usepackage{capt-of}
\usepackage{hyperref}
\usepackage{float}
\usepackage{lineno}
\usepackage{siunitx}
\usepackage{placeins}
\usepackage{subcaption}

\begin{document}

\begin{frontmatter}
\title{First 3D vector tracking of helium recoils for fast neutron measurements at SuperKEKB}

\author[addUH]{M. T. Hedges\corref{cor1}\fnref{purdue}}
\cortext[cor1]{Corresponding author: hedges7@purdue.edu}
\fntext[purdue]{Now at Purdue University, West Lafayette, IN 47907, USA}

\author[addUH]{S. E. Vahsen}
\author[addUH]{I. Jaegle\fnref{jnl}}
\fntext[jnl]{Now at Thomas Jefferson National Accelerator Facility, Newport News, VA, 23606, USA}
\author[addUH]{P. M. Lewis\fnref{bonn}}
\fntext[bonn]{Now at University of Bonn, Institute of Physics, Nußallee 12, 53115 Bonn, Germany}
\author[addKEK]{H. Nakayama}
\author[addUH]{J. Schueler}
\author[addUH]{T. N. Thorpe\fnref{ucla}}
\fntext[ucla]{Now at University of California Los Angeles, Los Angeles, CA 90095-1547, USA}

\address[addUH]{University of Hawaii, Department of Physics and Astronomy, 2505 Correa Road, Honolulu, HI 96822, USA}
\address[addKEK]{High Energy Accelerator Research Organization (KEK), Tsukuba, 305-0801 Japan}

\begin{abstract}
We present results from the first deployment of novel, high definition, compact gas Time Projection Chambers (TPCs) with pixel chip readout as part of the BEAST II beam background measurement project at SuperKEKB. The TPCs provide detailed 3D imaging of ionization from neutron-induced nuclear recoils in a helium and carbon dioxide target gas mixture at standard temperature and pressure. We present the TPC performance and the neutron backgrounds observed during the initial stage of collider commissioning. We find excellent electron background rejection, leading to background-free nuclear recoil measurements above 50~keV$_{\rm ee}$, despite the extreme high-background environment. We measure an angular resolution better than \(20^{\circ}\) for recoil tracks longer than 1.7~mm, corresponding to an average ionization energy of approximately 100 keV$_{\rm ee}$. We also obtain the full 3D vector direction of helium recoils by utilizing charge profile measurements along the recoil axis, with a correct head/tail assignment efficiency of approximately 80\%. With this performance, we present comparisons between measured and simulated event rates, recoil energy spectra, and directional distributions originating from beam-gas and Touschek beam losses at SuperKEKB. We utilize head/tail recognition to distinguish neutron components travelling with positive radial velocity in the Belle II coordinate system from those travelling in the opposite direction. Finally, we present a novel method of discriminating beam-gas interactions from Touschek beam losses that can eliminate the need for dedicated accelerator runs for background measurements. This method is still statistics-limited. However, future studies should be able to verify this method, which in turn could lead to neutron background analysis runs symbiotic with normal Belle II operation. The capabilities demonstrated here also suggest that high definition recoil imaging in gas TPCs is applicable to low energy, low-background experiments, such as directional dark matter searches.

\end{abstract}
\end{frontmatter}


\section{Introduction}
The BEAST II experiment provided measurements of beam-induced backgrounds to aid in early single-beam commissioning of the SuperKEKB accelerator and the Belle II experiment \cite{lewis19_first_measur_beam_backg_at_super}. The experimental campaign consisted of commissioning the SuperKEKB High Energy Ring (HER) and Low Energy Ring (LER) without final focusing required for beam collisions and without the Belle II detector installed in the interaction region. This provided an ideal environment to investigate effects leading to gradual reduction of current in the individual beams, or \emph{beam losses}, without the presence of luminosity-dependent and Belle II detector effects. This allows for validation of the methodology used for modeling beam losses, which in turn is used for estimating backgrounds during the Belle II physics campaign.

The expected beam loss mechanisms in these early conditions, as described in Ref. \cite{lewis19_first_measur_beam_backg_at_super}, are synchrotron radiation, the Touschek Effect \cite{Bernardini_1963}, beam-residual gas interactions (beam-gas) \cite{Moller99beam-residualgas}, and beam injection \cite{gabriel2020time}. Of these, the Touschek and beam-gas losses are of particular significance. Touschek losses occur when intra-beam Coulomb scattering causes particles to leave their nominal orbit. This effect scales with the charge-density of beam bunches. Beam-gas losses occur when particles in the beam collide with residual gas atoms in the sub-ideal vacuum of the beamline.

Given that SuperKEKB is an \(e^+e^-\) collider, the beam losses are expected to predominantly produce electromagnetic backgrounds. However, these beam losses also produce a relatively small but not insignificant amount of neutron backgrounds via electrofission and photofission \cite{osti_5872030,Baldwin_1947}. This neutron background is not only difficult to detect without dedicated instrumentation, but also difficult to model and simulate accurately. As such, neutrons have proven to be a particularly pernicious background at B-factories. In previous experiments, neutrons degraded the KLM detector performance at Belle \cite{neutrons_in_klm, Hoshi:2006ir} and the DIRC detector performance in the BaBar Experiment \cite{neutrons_in_dirc}. Given the beam current and luminosity upgrades at SuperKEKB, neutron backgrounds are expected to be a critical issue at Belle II. Further discussion on this topic can be found in Sec.~10.4 in the Belle II Technical Design Report \cite{Abe:1304162}.

The BEAST TPCs were designed to provide measurements of nuclear recoils from fast neutrons at SuperKEKB in order to test the validity of beam background simulations, with results presented in Ref.~\cite{hedges,lewis19_first_measur_beam_backg_at_super}. 
The BEAST TPCs constitute part of a broader University of Hawaii detector R\&D program that is also targeting Directional Dark Matter Detection. The dark matter component of this program was initially known as the Directional Dark Matter Detector (D$^3$) project~\cite{Vahsen:2011qx}, but has recently expanded and merged with a new, global effort called CYGNUS~\cite{Vahsen:2020pzb}.

Here we present final neutron background results from SuperKEKB Phase 1. We begin with a description of the TPC subsystem, followed by our energy calibration procedure. We then present and discuss their performance in detecting nuclear recoils during the first phase of SuperKEKB commissioning. Finally, we discuss the TPC performance and analyses in the context of other applications such as low-background directional dark matter detection experiments.

\begin{figure*}
\centering
\includegraphics[width=0.8\textwidth]{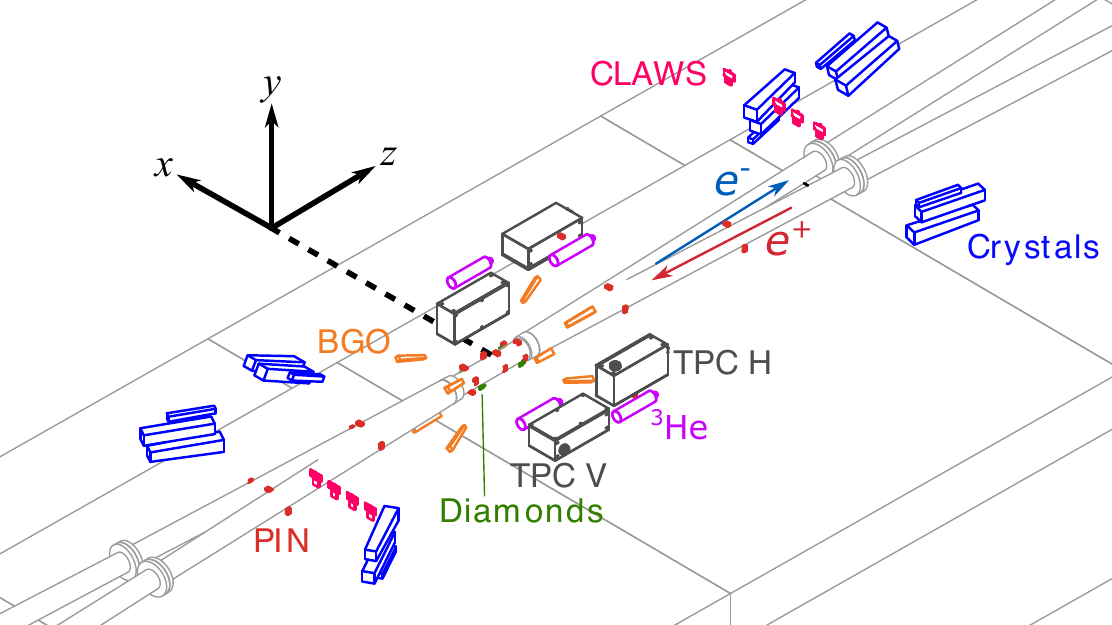}
\caption{Rendering of the full BEAST II Experiment with the TPCs shown in black, adapted from Ref. \cite{lewis19_first_measur_beam_backg_at_super}.}
\label{fig:beast} 
\end{figure*}

\section{TPC description and performance}
\label{sec:org2a2ad41}

Detailed specifications of the BEAST TPCs can be found in Ref.~\cite{Jaegle_2019}. To summarize, ten independent gas-filled time projection chambers were constructed, with two used in this work. The two TPCs presented here were installed in separate locations along the SuperKEKB beam-line, with one TPC installed in the horizontal plane of the beam-orbit---named "TPC H"---and the other installed in the vertical plane of the beam-orbit---named "TPC V", with electronics and gas flow systems originating from a control room outside of the SuperKEKB beamline. The TPCs are designed to detect ionization from recoiling nuclei resulting from fast neutrons scattering with the target-gas mixture of \(^4_2\mathrm{He}\) and CO\textsubscript{2} (70\% He, 30\% CO\textsubscript{2}) inside of a 2.0 \texttimes{} 1.68 \texttimes{} 10.87 cm\textsuperscript{3} active volume contained within a larger vacuum vessel.

The TPCs detect neutrons as follows: approximately one in $10^4$ incident fast neutrons scatters elastically with a target atom in the gas mixture, producing a recoiling \(^4_2\mathrm{He}\), \(^{12}_6\mathrm{C}\), or \(^{16}_8\mathrm{O}\) nucleus. The recoil nucleus leaves a cloud of electron-ion pairs behind as it propagates inside the gas volume. An applied electric field of 530 V/cm inside the gas volume causes the electrons from the ion pairs to drift at a constant drift velocity of 10 \si\micro m/ns, as calculated by Magboltz \cite{magboltz}, through two sequential Gas Electron Multipliers (GEMs) \cite{Sauli_1997}. A high voltage applied to each GEM results in a gain with magnitude between 10–100 per GEM. This lower-gain mode was chosen in order to prioritize long-term stability in performance over sensitivity.

The He:CO$_2$ gas mixture meets a number of safety, gas detector, and neutron detection requirements. First, all gas components are non-toxic and non-flammable, which was an important consideration given the unpredictable commissioning environment and planned operation in underground and confined spaces. Second, the mixture provides high gas avalanche gain and good operational stability. Third, helium provides near maximal energy exchange in elastic scattering with neutrons. For the fast-neutron energies of interest, this results in recoils short enough to be fully contained in the compact detectors yet long enough for the recoil direction to be clearly discernible after diffusion. The detailed mixture optimization is described further in Ref. \cite{Jaegle_2019}.

The amplified charge is detected by a custom-metalized ATLAS FE-I4B pixel ASIC which interfaces with the USBPix \cite{Backhaus_2011} and SEABAS \cite{Uchida_2006} data acquisition systems. Ref.~\cite{Garcia_Sciveres_2011} provides detailed documentation on the design and performance of the pixel chip. To summarize, the chip is an array of pixels arranged in 80 columns \texttimes{} 336 rows. Each pixel has dimensions of \(250 \times 50\) \si\micro m\textsuperscript{2}, resulting in a 2 \texttimes{} 1.68 cm\textsuperscript{2} active area for the entire array. The columns and rows define the \(x\) and \(y\) coordinates of the TPC internal coordinate system, respectively. The system operates in a self-triggered mode where a trigger occurs once any pixel records charge above a threshold of 2600 electrons. An event starts with a triggered pixel and lasts until either 100 clock cycles (2500 ns) have elapsed or all pixels return under threshold. The chip records Time-Over-Threshold (TOT) and the number of clock cycles since the beginning of the trigger for all pixels in the event. The number of clock cycles in addition to the constant electron drift velocity can be used to infer a \(z\) coordinate relative to other pixels in the event. The pixels integrate charge drifting down from the GEMs, thus the inferred \(z\) coordinate corresponds to the time in which the integrated charge rises above the pixel threshold. Given that the TPCs are self-triggered, all z-coordinates are initially relative. We have previously demonstrated that our high readout segmentation also enables measurement of the absolute z-coordinate of events via the transverse diffusion of drift charge \cite{Lewis:2014poa}. However, we do not use this capability here, meaning that all \(z\) coordinates presented in this paper are relative. Absolute \(z\) serves as a rejection of backgrounds that originate from inside the detector volume, which is required in low background experiments. The backgrounds relevant here come from the SuperKEKB accelerator, and are thus external to the detector volume.

TPCs with GEM amplification and pixel chip readout were chosen as the preferred detector technology for the BEAST II experiment because of their flexibility, robustness, and compact size. A high resolution gas TPC is flexible in that it can reconstruct the direction of, and clearly distinguish between charged tracks, electron recoils from gamma rays scattering, and nuclear recoils from neutron scattering. Given the large uncertainties in a commissioning environment, we felt this would allow us to investigate not only neutrons, but also other background components, if the need should arise. The double GEM configuration is robust against ion feedback, and thereby improves the maximum rate capability of the detectors, compared to wire-based TPC readout. The relatively low gamma-ray and neutron interaction efficiency of the gas target effectively provides a prescale, further increasing the ability of these detectors to tolerate a high-rate environment. The final consideration was available space. As outlined in Ref. [14], each TPC was required to fit within a $14 \times 16 \times 40~\rm{cm}^3$ volume. Alternative directional neutron detector technologies available at the time, such as compact neutron scatter cameras \cite{neutron-scatter-camera} were considered impractical due to their size.  The main concern about our technology choice was the damage of pixel chips in the case of accidental electrical discharges from high voltage components in the detectors. However, such damage has not occurred throughout many years of operation. Following the work described here, eight detectors were used to measure neutron backgrounds inside the Belle II detector \cite{zac-phase2}, and then six of these were repurposed for measuring the neutron flux in the SuperKEKB tunnel \cite{jeff-phase3}. These detectors are still operating stably and providing real-time neutron rates to the SuperKEKB control room at the time of writing.

\subsection{Track reconstruction}
\label{sec:orgb3120f8}

The reconstructed 3D coordinates from the activated pixels are fit to form a track. The fit algorithm is a MINUIT \cite{James:2296388}-based \(\chi\)\textsuperscript{2} minimization of a straight 3-dimensional line. The algorithm returns five parameters: polar and azimuthal angles---\(\theta\) and \(\phi\), respectively---and the $x$, $y$, $z$ coordinates of an arbitrary point along the line. The location of the TPCs in the Belle II coordinate system are shown in Fig \ref{fig:beast}. Of the five fit parameters, we are most interested in \(\theta\) and \(\phi\), as they provide the directional information of the reconstructed track. Angular resolution is limited by diffusion and multiple elastic scattering of the recoiling nucleus in the gas. This can be seen in Fig. 6, which shows a typical nuclear recoil signal event that does not travel along a straight line. Here, we do not attempt to separate multiple scattering from detector resolution.

\subsection{Charge recovery and energy scale calibration}
\label{sec:eCal}
The pixels measure integrated charge in units of TOT, which must be calibrated in order to infer information about the primary charge produced by the nuclear recoil. This procedure consists of: (a) calibrating the pixel chip response, (b) recovering charge lost above the pixel saturation point, (c) recovering charge lost below the pixel threshold, and (d) measuring and calibrating the effective GEM gain.

\subsubsection{Calibration of pixel chip charge scale}
\label{sec:fei4-cal}

Individual pixel calibrations are performed via a test-pulser on the chip \cite{Garcia_Sciveres_2011}. Test pulses of varying charge values  are repeatedly injected into each pixel as the DAQ software fine-tunes the charge-threshold and charge-integration time iteratively until variation in performance of all pixels is minimized at the desired threshold setting. We choose a threshold of 2600 electrons. This calibration also measures the mean of the pixel noise to be of the order  of a few hundred electrons. Since the noise level is approximately an order of magnitude below the threshold, the TPCs can achieve steady-state operational running without noise triggering the readout. This calibration process allows converting units of TOT in individual pixels into detected electrons via the value of the capacitance of the built in test-pulser. While individual pixel calibrations can be applied, the spread in performance across individual pixels is small enough that the mean value is applied to all pixels in the event.

The TOT value for a given pixel can be any integer number from 0 through 13, where $\mathrm{TOT} = 0$ is our chosen threshold of 2600 electrons, and $\mathrm{TOT} = 13$ corresponds to the maximum detectable charge that a pixel can resolve, which is determined by the chosen configuration for the chip. We define pixels with $\mathrm{TOT} = 13$ to be \emph{saturated}. By definition, an individual pixel cannot provide information for integrated charge below its threshold or above its saturation point.


\subsubsection{Charge recovery from pixel saturation}

We aim to recover charge lost from these effects through a charge recovery analysis developed using Monte Carlo data. We utilize a full Geant4 detector simulation of nuclear recoil events in the TPC that includes the charge-loss effects from the threshold and saturation limitations of the pixel chip  \cite{lewis19_first_measur_beam_backg_at_super}. We then compare the total detected charge to the truth recoil energy for fully reconstructible events. We select fully reconstructible nuclear recoil events by implementing a fiducialization selection where we exclude events with pixels with hits within 500 \si\micro m of the edge of the pixel chip.

\begin{figure}[htbp]
\centering
\includegraphics[width=.9\linewidth]{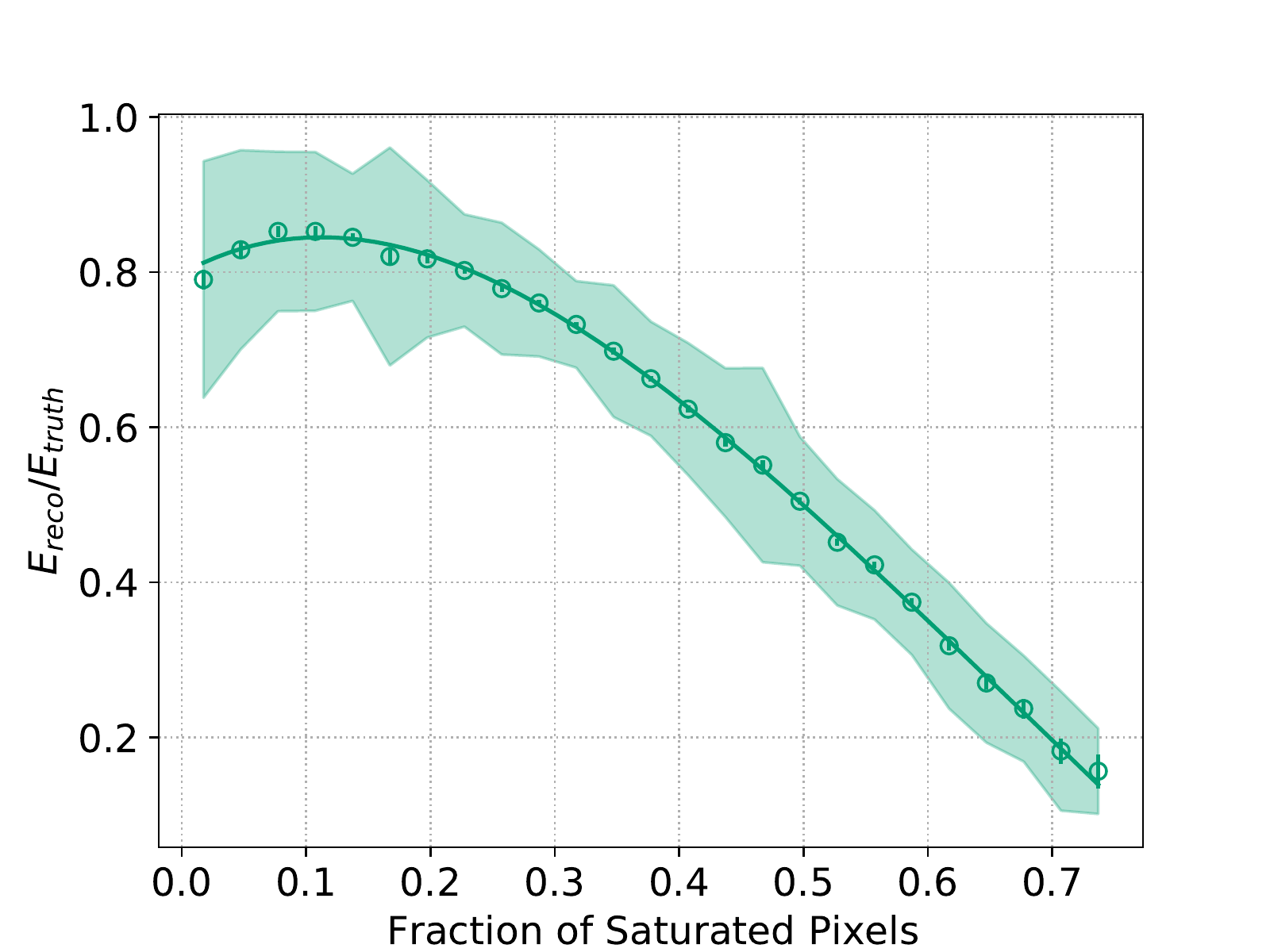}
\caption{\label{fig:sat-pix}Ratio of reconstructed to true energy versus fraction of saturated pixels per event in all simulated recoils fit to a fourth-order polynomial. The shaded area represents one standard deviation of the data in each bin.}
\end{figure}

After applying the fiducialization, we obtain the simulated detected charge in the event by converting the total TOT from all pixels in the event into charge according to the pixel chip configuration. We then calculate the detected ionization energy, $E_{\mathrm{reco}}$, by accounting for the simulated, idealized gain of 1500 and the gas work function, $W = 35$ eV/ion-pair, as calculated by Garfield++ \cite{garfield}. We then plot the ratio of \(E_{\text{reco}}\) to the truth recoil energy, \(E_{\text{truth}}\), versus the number of saturated pixels in the event. The points in Fig. \ref{fig:sat-pix} show the mean and error on the mean of all simulated recoil events in each bin and the shaded region shows one standard deviation of each bin. We fit a fourth-order polynomial to the data to obtain a correction function for the charge lost due to pixel saturation. 

\subsubsection{Charge recovery from pixel threshold}

\begin{figure}[htbp]
\centering
\includegraphics[width=.9\linewidth]{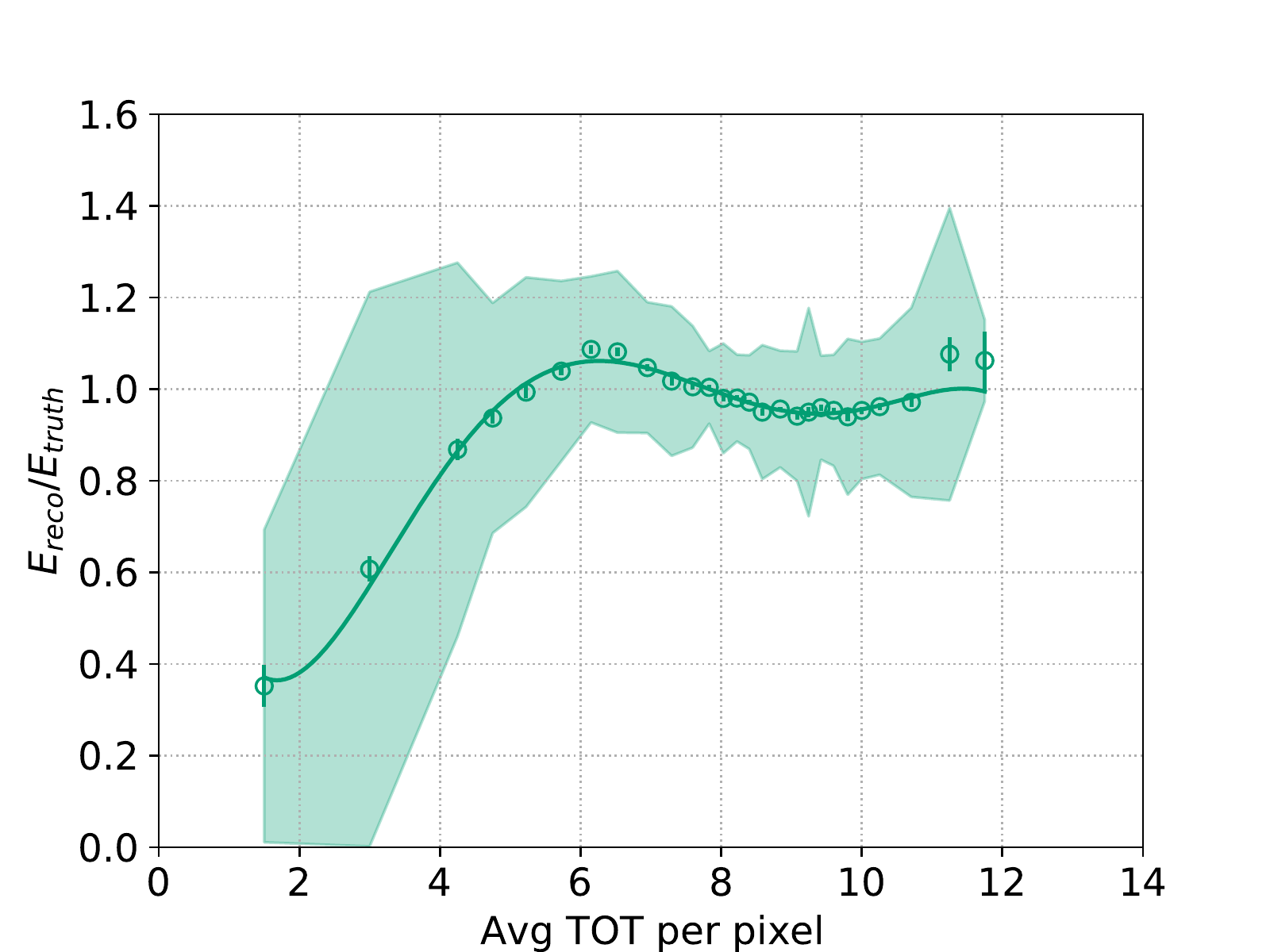}
\caption{\label{fig:thresh-pix}Ratio of reconstructed to true energy versus the average TOT in a single pixel per event fit to a fifth order polynomial.
The shaded area represents one standard deviation of the data in each bin.}
\end{figure}

We perform a similar procedure to account for charge lost due to the pixel threshold. We can see this by plotting \(E_{\text{reco}}/E_{\text{truth}}\) versus the average TOT in the event after correcting for charge-loss from saturation. We bin and fit this distribution to a fifth-order polynomial. The distribution and the corresponding fit function are shown in Fig. \ref{fig:thresh-pix}. 

\begin{figure*}[hbt]
\centering
\includegraphics[angle=90,width=0.8\textwidth]{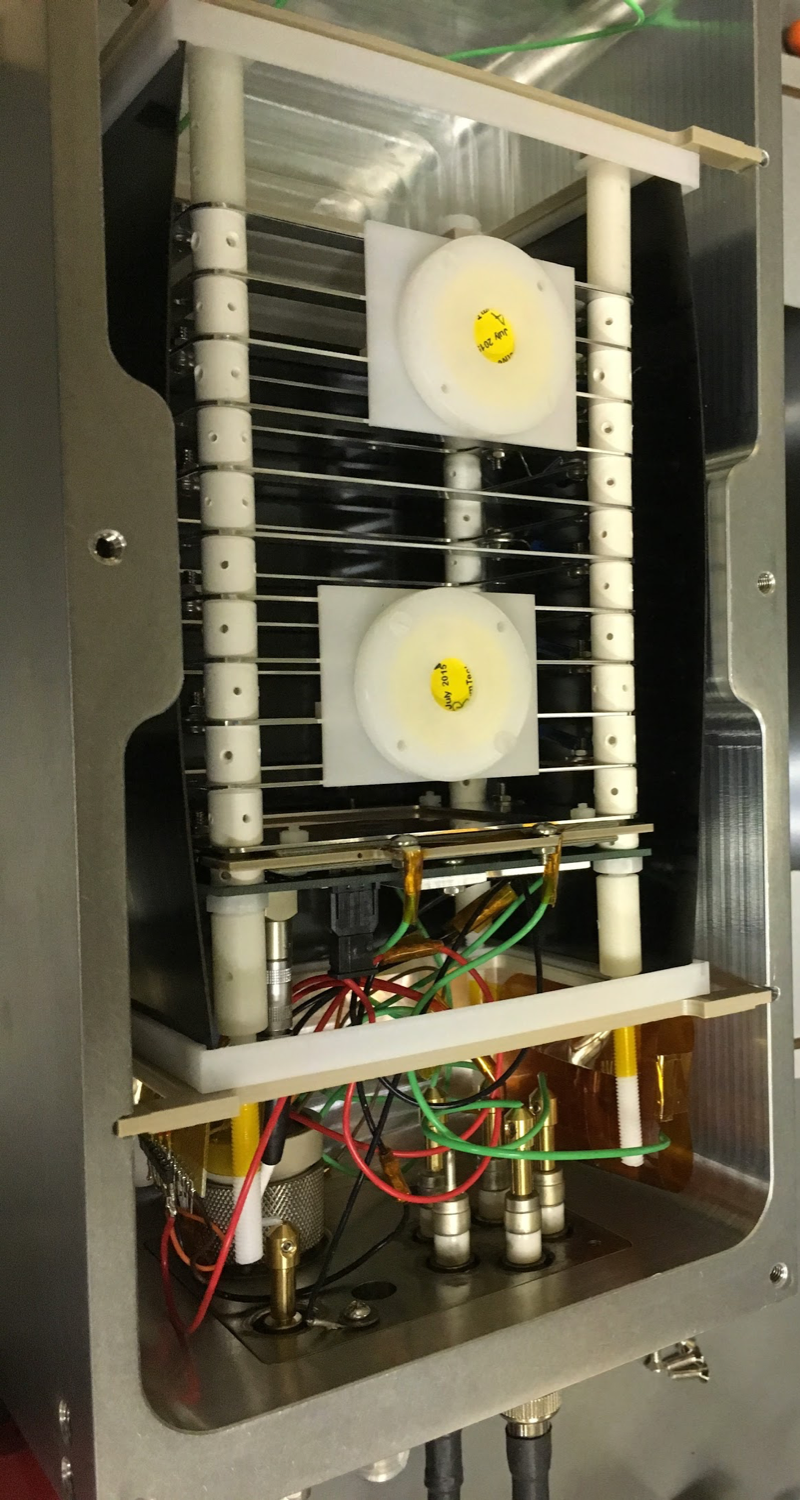}
\caption{A photo of the inside of a TPC showing the $^{210}$Po calibration sources. The white containers with the yellow centers hold the sources. The container to left of the photo is the “top” source---the source at larger drift distance---and the container to the right of the photo, closest to the green wires, is the “bottom” source. Adapted from Ref. \cite{lewis19_first_measur_beam_backg_at_super}.}
\label{fig:tpc-internals}
\end{figure*}

\subsubsection{GEM gain calibration}

The sum of all charge collected by the pixel chip after correcting for charge-loss effects results in a measurement of the total charge post-amplification. To convert this charge back into primary ionization generated by the recoil, we must measure the double-GEM effective gain. Gas detector gain can specifically be both time and event rate dependent. In TPCs, different drift lengths can additionally produce varying amounts of charge below threshold. Given the large expected variations in event rate during SuperKEKB commissioning, we decided to utilize two calibration sources inside each TPC vessel to continuously monitor the effective gain in-situ for two different drift lengths.

The sources used are 10~nCi \(^{210}\mathrm{Po}\) sources that emit alpha particles with average energy of \(\sim\)5 MeV. The two sources are installed on the outside of the field-cage at different drift lengths and along different pixel columns, thereby allowing for discrimination of each source individually. The source installed at largest drift distance is referred to as the “top” source, and the other is referred to as the “bottom” source. The physical setup of this configuration is shown in Fig. \ref{fig:tpc-internals}.

\begin{figure}
\includegraphics[width=\columnwidth]{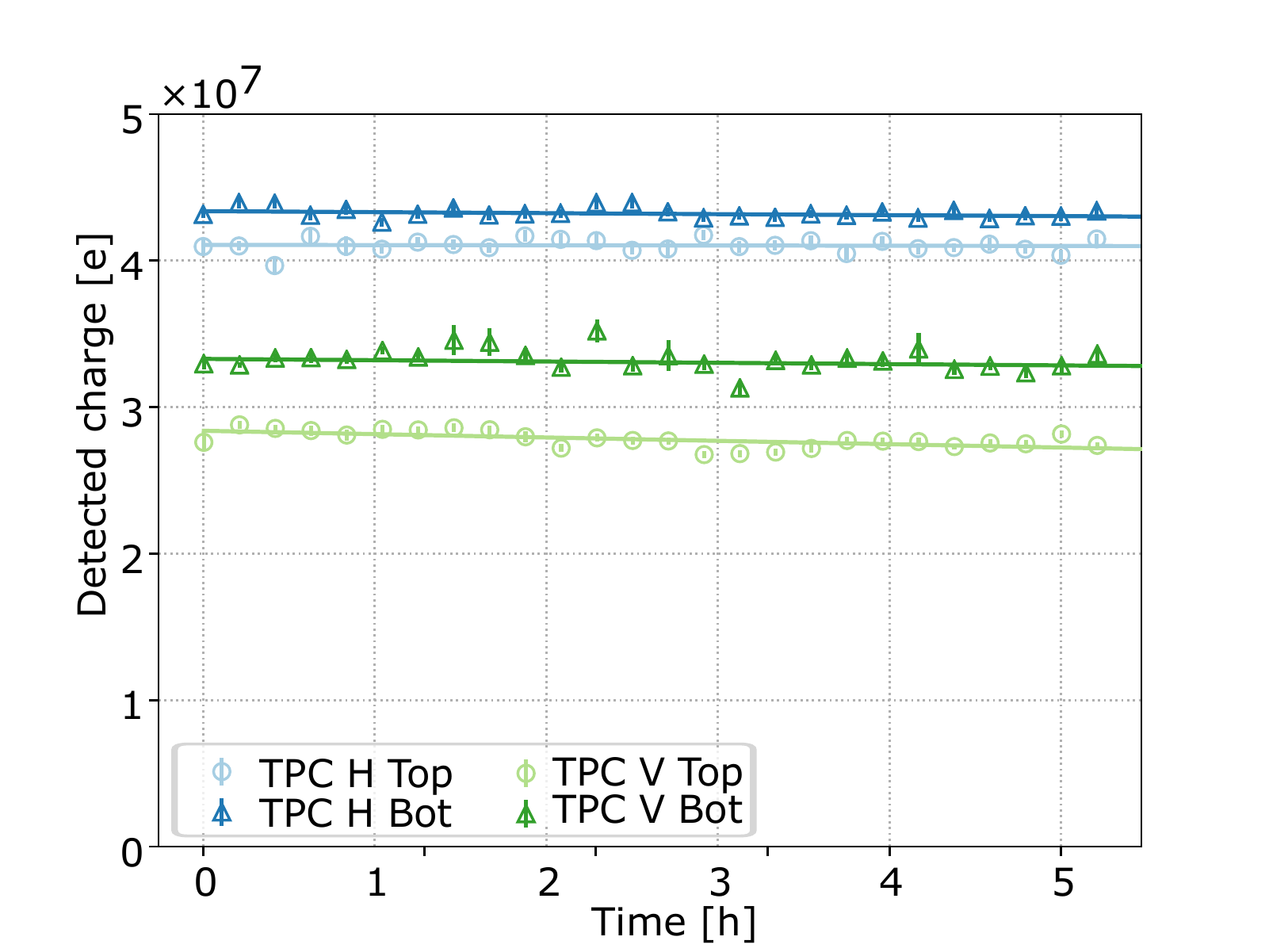}
\caption{Detected charge of alpha particle calibration events in TPC H in blue and TPC V in green versus time.}
\label{fig:tpc-gain-stability}
\end{figure}

\begin{figure}[htbp]
\centering
\includegraphics[width=.9\linewidth]{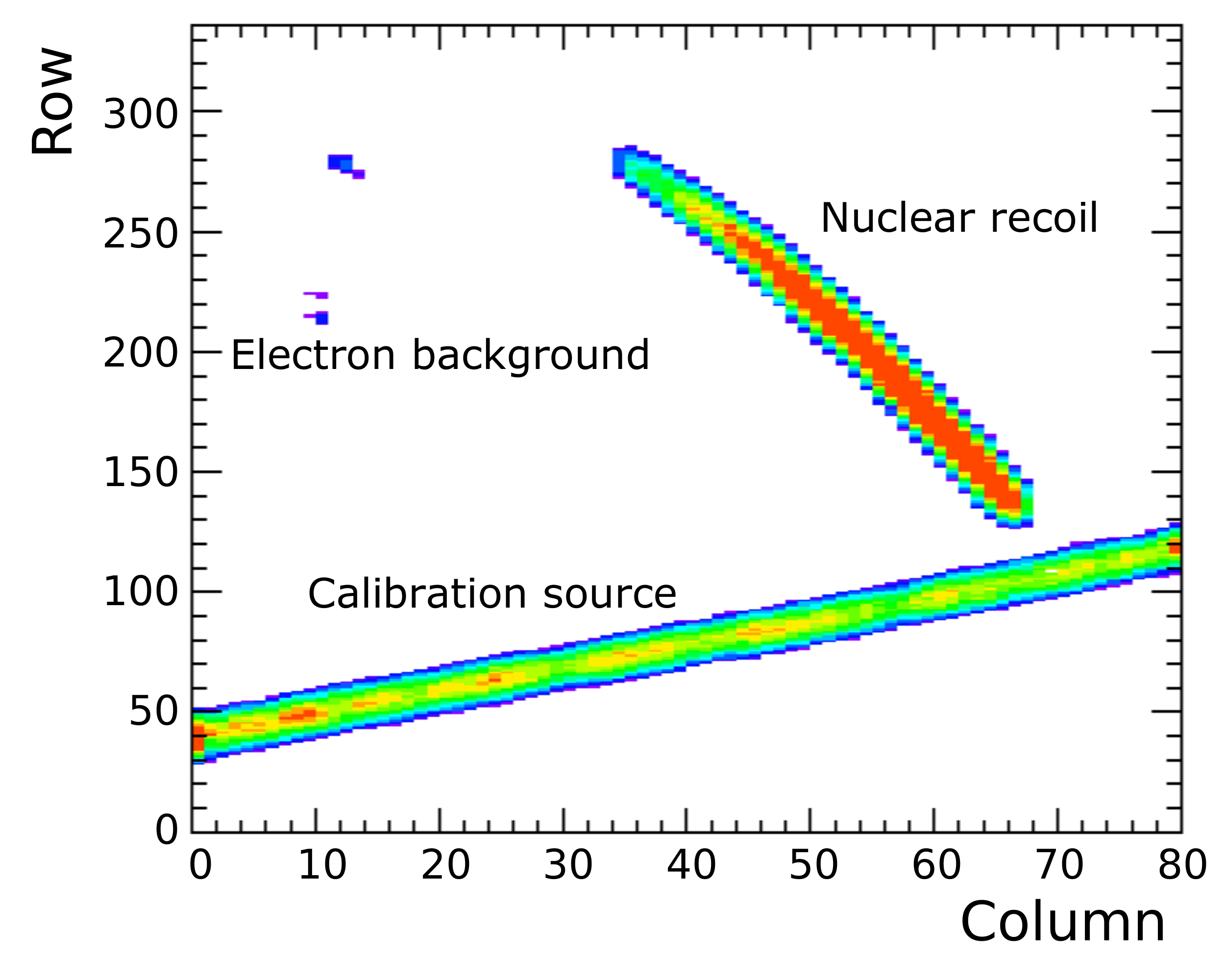}
\caption{\label{fig:evt-display}Adapted from Ref. \cite{lewis19_first_measur_beam_backg_at_super}. Three separate events, illustrating the typical signatures of electron recoils, alpha particles from the internal calibration sources, and the nuclear recoil signal. The horizontal and vertical axes show the row and column number, respectively, of the pixels on the FE-I4B pixel chip. The color illustrates the amount of charge detected in each pixel.}
\end{figure}

After applying the previous charge-recovery correction, we next obtain a relative gain correction factor for each TPC by comparing in-situ measurements of events from the calibration sources to a dedicated Geant4 Monte Carlo simulation. To achieve this, we first check the stability of the gain versus time. This requires selecting a sample of events from the calibration sources in a TPC. Reconstructed alpha events are selected by their unique signal of a track with a large \(\mathrm{d}E/\mathrm{d}x\) that is long enough to span the entire width of the pixel chip, as shown in Fig. \ref{fig:evt-display}.

We check the stability of the signal from these events by plotting the total detected charge per event versus time. This is shown on our two detectors, TPC H and TPC V, in Fig. \ref{fig:tpc-gain-stability}. We note from this plot that the gain is very stable over the course of many hours. Specifically, the slope of each line is very near zero, indicating fluctuations in stability on the order of a few percent per hour, at most, over the course of nearly 6 hours. This justifies using time-independent calibrations.

Next, we calculate a gain correction factor for each TPC by comparing measured performance to an idealized performance obtained from simulated data. The first step is to determine the effects of charge lost due to the finite pixel charge scale dynamic range by plotting the total detected charge per unit of length, or \(\mathrm{d}Q/\mathrm{d}x\), of events from the calibration sources separately. We select candidate events in experimental and Monte Carlo data such that \(89.5^{\circ} < \theta < 90.5^{\circ}\), corresponding to \(\pm0.5^{\circ}\) from a perfectly horizontal track. This produces a sample of events similar to the example "Calibration source" event in Fig. \ref{fig:evt-display}.

The selected events should show two distinct peaks in the \(\mathrm{d}Q/\mathrm{d}x\) distributions for each source. We use these peaks to obtain correction factors for the gain of the physical TPCs to match the effective gain of the simulated TPC. To do this, we calculate an average \(\mathrm{d}Q/\mathrm{d}x\) for the two peaks for a given TPC. This provides one measurement of \(\mathrm{d}Q/\mathrm{d}x\) for calibration events drifting a length halfway between the two calibration sources. The calibration coefficient for each physical TPC is the ratio of its average \(\mathrm{d}Q/\mathrm{d}x\) to the corresponding value in simulated data. We then use this factor as a multiplicative correction for the total detected charge in every event.  We find that both TPCs have a lower effective gain than the idealized performance shown in the simulation, with TPC V needing a 57\% correction and TPC H needing a 22\% correction. 

\begin{figure}[htbp]
\centering
\includegraphics[width=.9\linewidth]{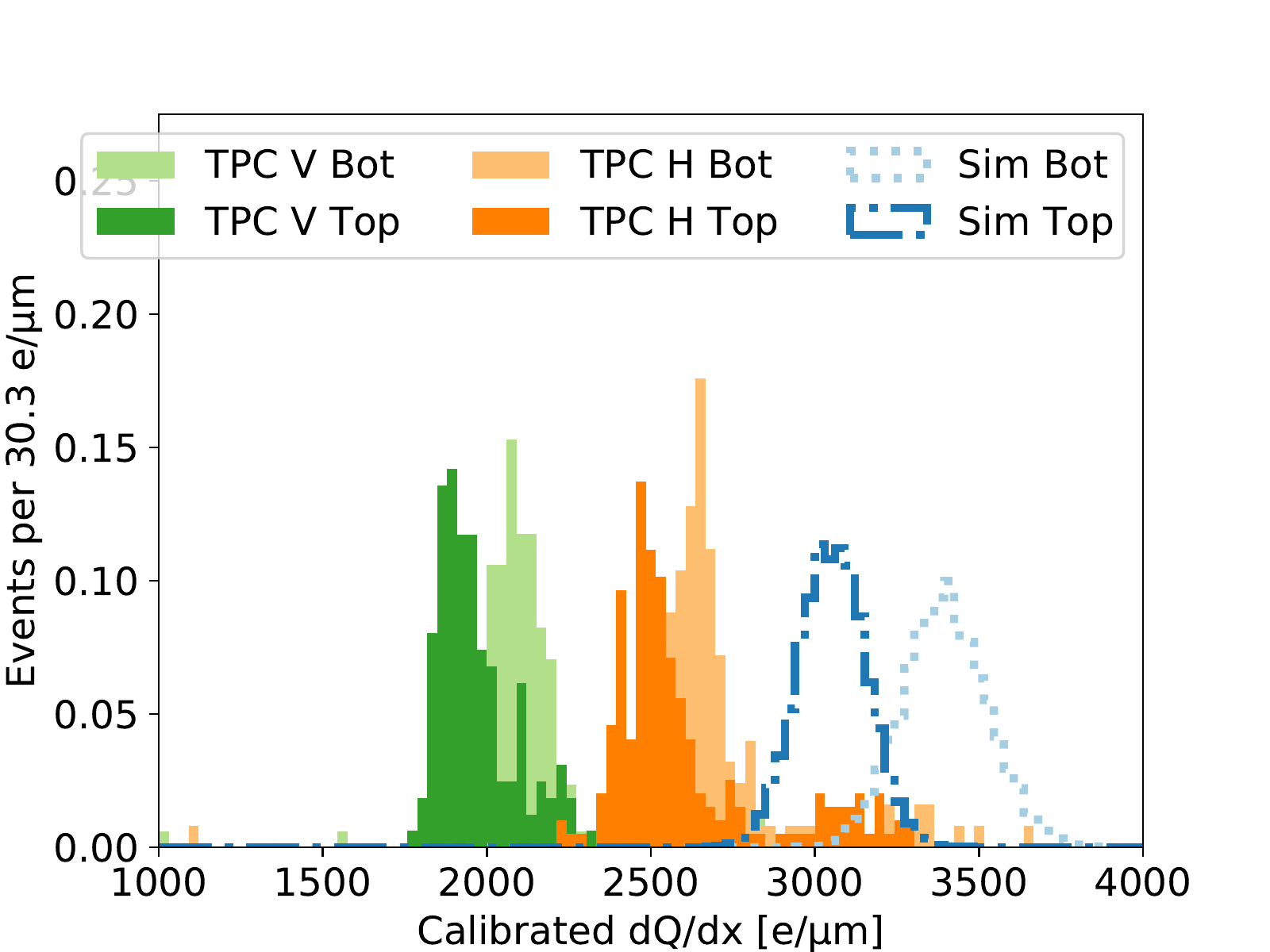}
\caption{\label{fig:dQdx-calibrated}Normalized histograms of the reconstructed detected charge divided by track length after correcting for pixel saturation and charge below the pixel threshold, via the fitted functions shown in Fig. \ref{fig:sat-pix} and \ref{fig:thresh-pix}.}
\end{figure}



\subsection{Background rejection}
\label{sec:orgc7b5f3d}
\label{sec:sels}

The following selections are applied to obtain a signal event sample for all analyses: (a) a firmware veto implemented at the trigger level rejects electron recoil events produced by the intense X-ray backgrounds at SuperKEKB, (b) a fiducial volume “edge veto,” which requires no pixels triggered within 500 \si\micro{}m of the four outer edges of the pixel chip ensures the drifted charge is entirely contained within the pixel chip area, (c) the fraction of saturated pixels and average TOT per pixel in the event falls within the coverage of the fits in Fig. \ref{fig:sat-pix} and \ref{fig:thresh-pix}, (d) the fitting algorithm used to fit the event to a straight line must converge, and (e) the ratio of calibrated detected energy to track length (\(\mathrm{d}E/\mathrm{d}x\)) is greater than 40 keV$_{\rm ee}$/mm, which removes electron recoil events and events from minimum-ionizing particles. We test these selections with a sample of 13,011 simulated events consisting of a signal sample---helium, carbon, and oxygen nuclear recoils---as well as a background sample---electrons, positrons photons and protons. This combined sample comes from simulation of the SuperKEKB beam-backgrounds in a full Geant4 model of the TPCs, which is described in Sec. \ref{sec:org45d4f85}.

The firmware veto in selection (a) is designed to limit triggering on events from
electron and X-ray backgrounds, such as those shown in the upper left of Fig. \ref{fig:evt-display}. The high rate of such events can saturate the data acquisition system and can lead to significant detector dead-time. We discriminate these events via their difference in \emph{trigger length} from the nuclear recoil signal. The trigger length of an event corresponds to the total length of time from when the integrated charge in any pixel crosses threshold until all pixels return under threshold. Since the trigger length is related to the total amount of detected charge, an electron event will have significantly shorter trigger length than a nuclear recoil. The veto is configured to reject events with a trigger length smaller than a configurable threshold.

The edge veto in selection (b) is used for two purposes. First, as can be seen Fig. 6, this selection separates nuclear recoils from alpha particles emitted by the calibration sources. Second, applying this fiducialization allows for selecting events where the  entire recoil is contained within the active area of the pixel chip.

For the \(\mathrm{d}E/\mathrm{d}x\) in selection (e), we convert the total detected charge back to primary ionization energy and divide by the length of the track. We use a corrected length, defined as \(L_C = L_{\mathrm{raw}} - w\) where \(L_{\mathrm{raw}}\) is the uncorrected 3D length of the track and \(w\) is the width of the track. \(L_{\mathrm{raw}}\) is calculated by projecting the pixel coordinates along the fitted track axis and returning the 3D distance between the two points of largest mutual projected-distance. The width \(w\) is defined such that 
$$w = \mathrm{max} \{\vec{p_i} \cdot (\hat{u} \times \hat{z})\} - \mathrm{min} \{\vec{p_i} \cdot (\hat{u} \times \hat{z})\}$$
where $\vec{p_i}$ is the vector of the $i^{\mathrm{th}}$ pixel hit in the event, $\hat{u}$ is the unit vector along the axis of the reconstructed track and $\hat{z}$ is the $z$-axis. The value of \(w\) gives an indication of the effect of diffusion of the track, thus the subtraction of \(w\) from \(L_{\mathrm{raw}}\) provides a more accurate representation of the track length.

\begin{figure}
\centering
\includegraphics[width=0.96\columnwidth]{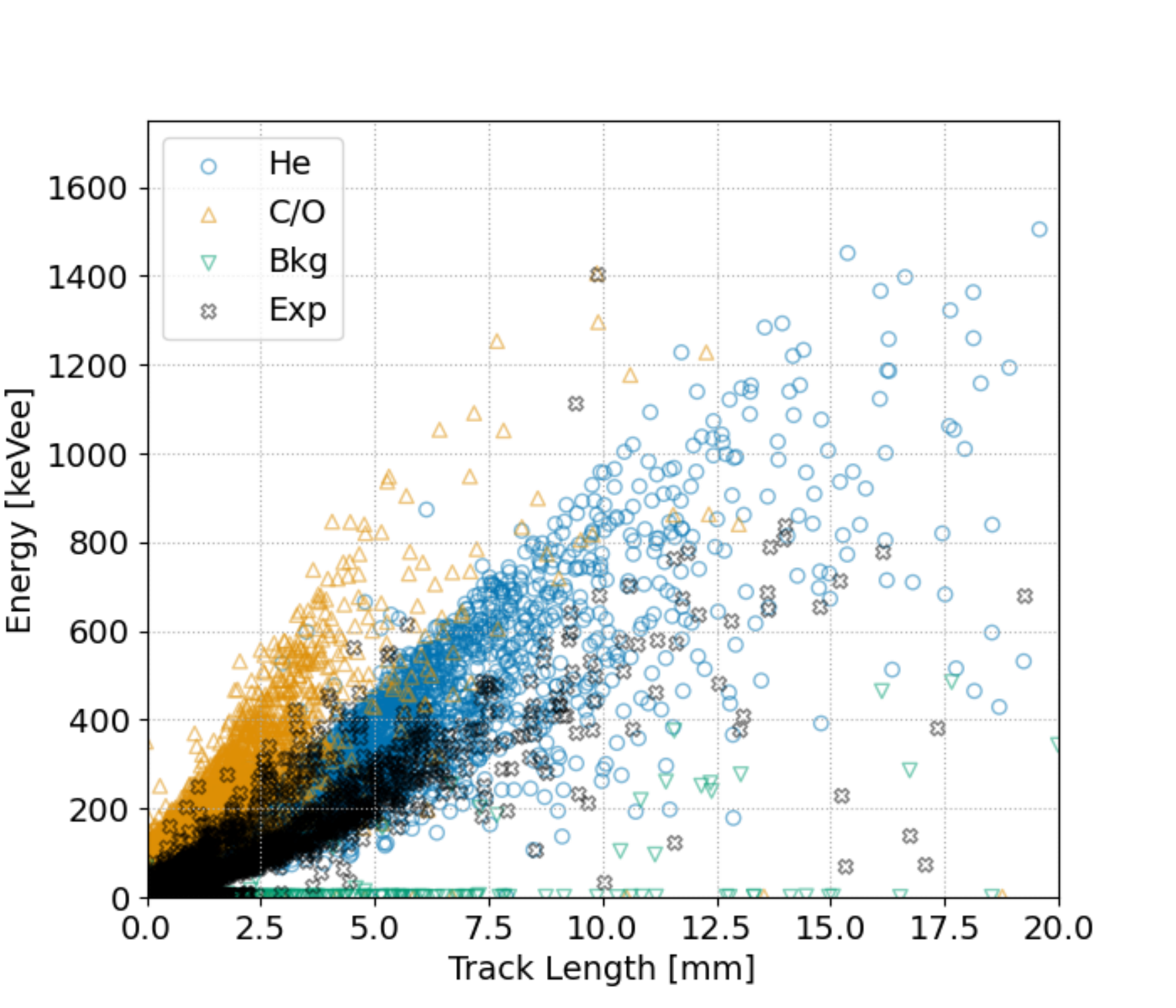}
\includegraphics[width=0.96\columnwidth]{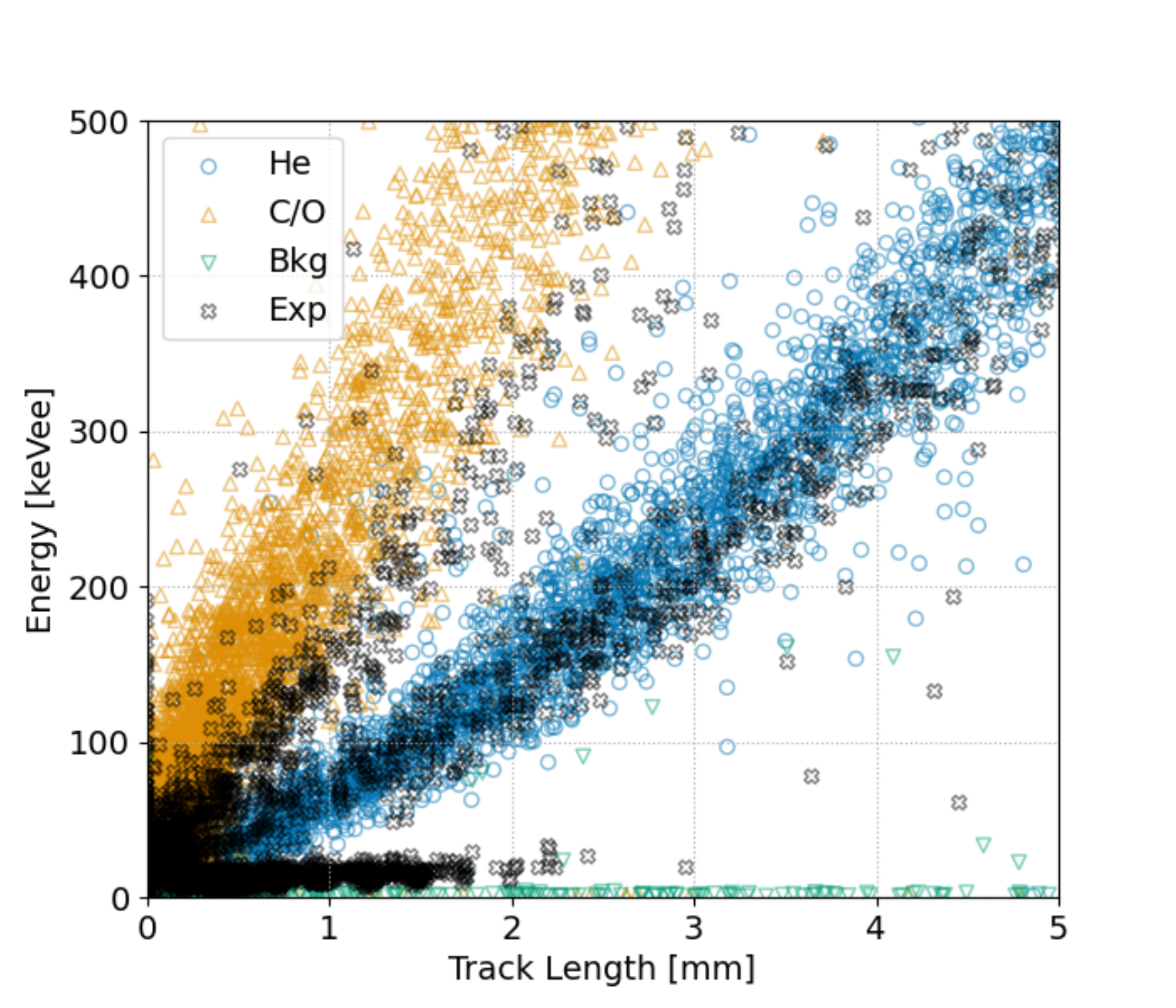}
\includegraphics[width=0.96\columnwidth]{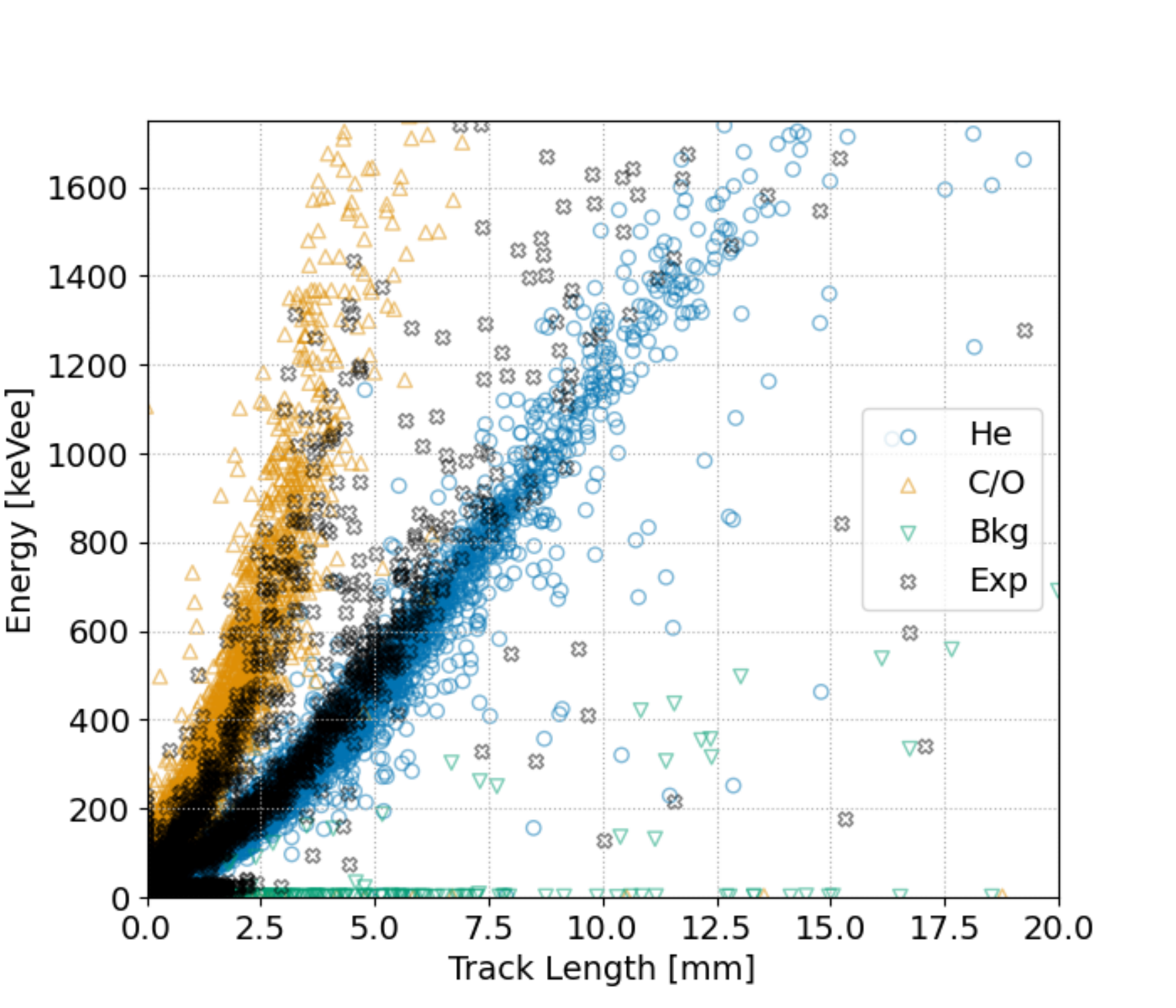}
\caption{Reconstructed energy versus reconstructed length in simulated and experimental data in both TPCs before charge-loss corrections (top), after charge-loss corrections (middle), and after an additional energy scale correction factor of 1.2 to experimental data (bottom).}
\label{fig:tpc-dEdx}
\end{figure}

After applying these selections, we plot \(\mathrm{d}E/\mathrm{d}x\) for Monte Carlo and experimental data. For comparison, we first show reconstructed energy versus length by only applying a relative gain correction, i.e. without applying the results of the charge recovery analysis of the previous section. This is shown in the top distribution in Fig. \ref{fig:tpc-dEdx}. It can be seen here that there is a clear separation of signal recoils from background events, which consist of electrons, positrons, and protons, as well as a visible separation of recoils from helium and recoils from carbon and oxygen. However, there is visible disagreement between simulated and experimental data for the helium, carbon, and oxygen recoils.

The middle plot shows the effect of including the results of the charge recovery analysis. While the agreement between simulated and experimental data is undoubtedly better, there is still a noticeable difference. Specifically, we see that the amount of detected energy in simulated data is overestimated along both the helium band and the carbon and oxygen band as the recoil energy increases. We find that an additional increase of 20\% in the detected energy in experimental data results in better agreement between simulated and experimental helium events.

Applying this 20\% increase to events in experimental data is shown in in the bottom distribution of Fig. \ref{fig:tpc-dEdx}, where it is immediately apparent that agreement between the Monte Carlo and experimental helium bands is better, but not perfect. While the agreement in helium improves with this factor, there is still noticeable disagreement in the carbon and oxygen band, where the detected recoil energy is higher by as much as $\sim$~50\% in simulated data. This discrepancy may be due to modeling of the primary ionization in simulation, as a single correction factor cannot be used for C, O, and He recoils.\footnote{We accept this discrepancy in modeling He from C/O recoils for the scope of our fast neutron studies, where we are primarily focused on selecting helium recoils. However, we recommend this effect be studied further, ideally with smaller-scale, dedicated TPC simulation and experimental data outside the context of simulating and analyzing accelerator backgrounds.}

From these figures, it can be seen that separating our nuclear recoil signal from background should be possible based on a selection of \(\mathrm{d}E/\mathrm{d}x\). In order to optimize this selection, we look specifically at the efficiency, \(\epsilon\), and purity, \(p\), of selecting events based on a given value of \(\mathrm{d}E/\mathrm{d}x\) in the Monte Carlo sample. Here, \(\epsilon\) is defined such that \(\epsilon = N_{\mathrm{sel}}/N_{\mathrm{sig}}\), where \(N_{\mathrm{sel}}\) is the number of signal events passing selections , and \(N_{\mathrm{sig}}\) is the total number of signal events in the Monte Carlo sample. The purity is defined such that \(p = N_{\mathrm{sel}} / (N^\prime_{\mathrm{sig}} + N^\prime_{\mathrm{bkg}}) \), where \(N^\prime_{\mathrm{sel}}\) is the number of selected signal events and \(N^\prime_{\mathrm{bkg}}\) is the amount of selected background events in the Monte Carlo sample.

\begin{figure}
\includegraphics[width=\columnwidth]{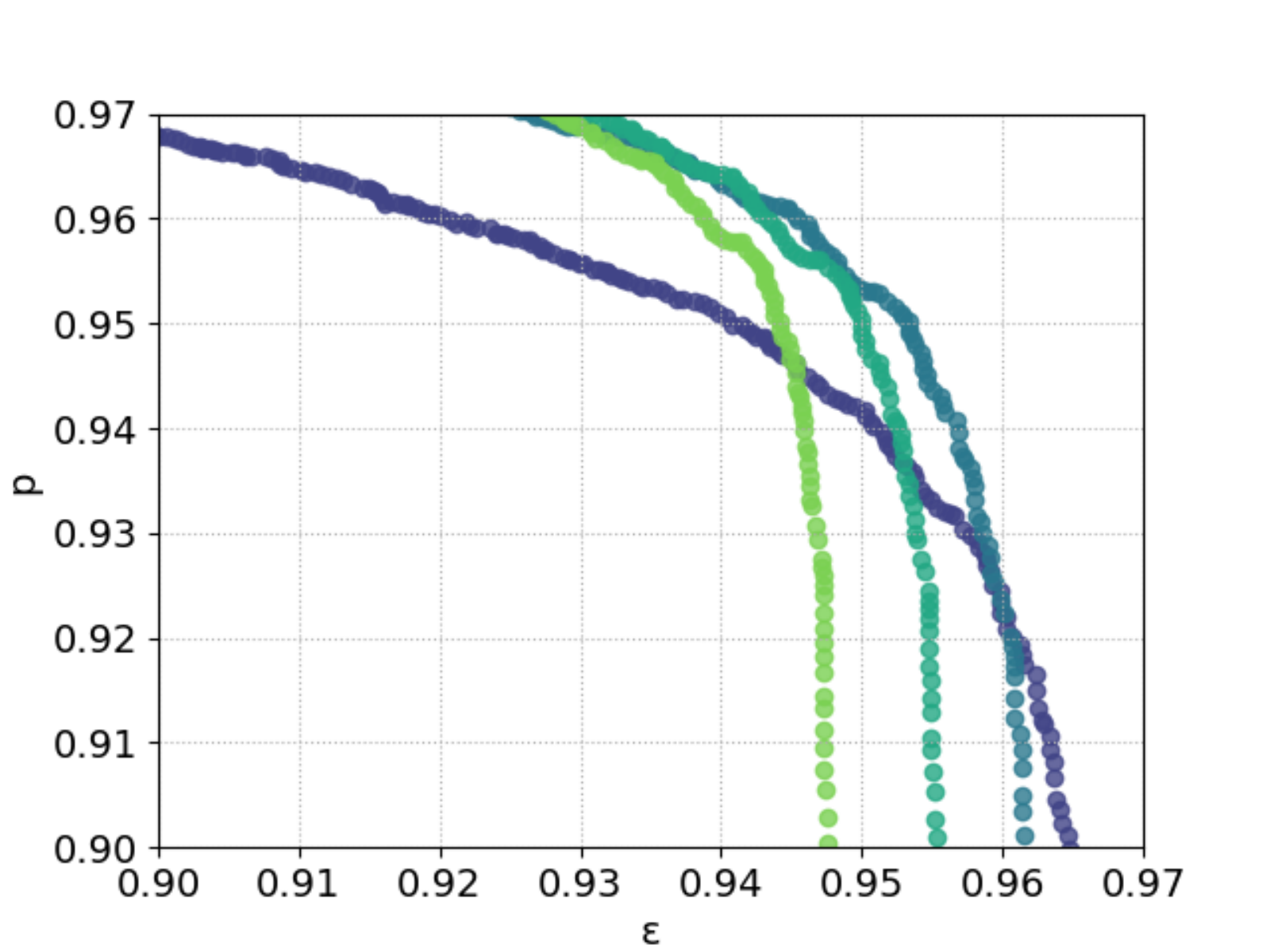}
\includegraphics[width=\columnwidth]{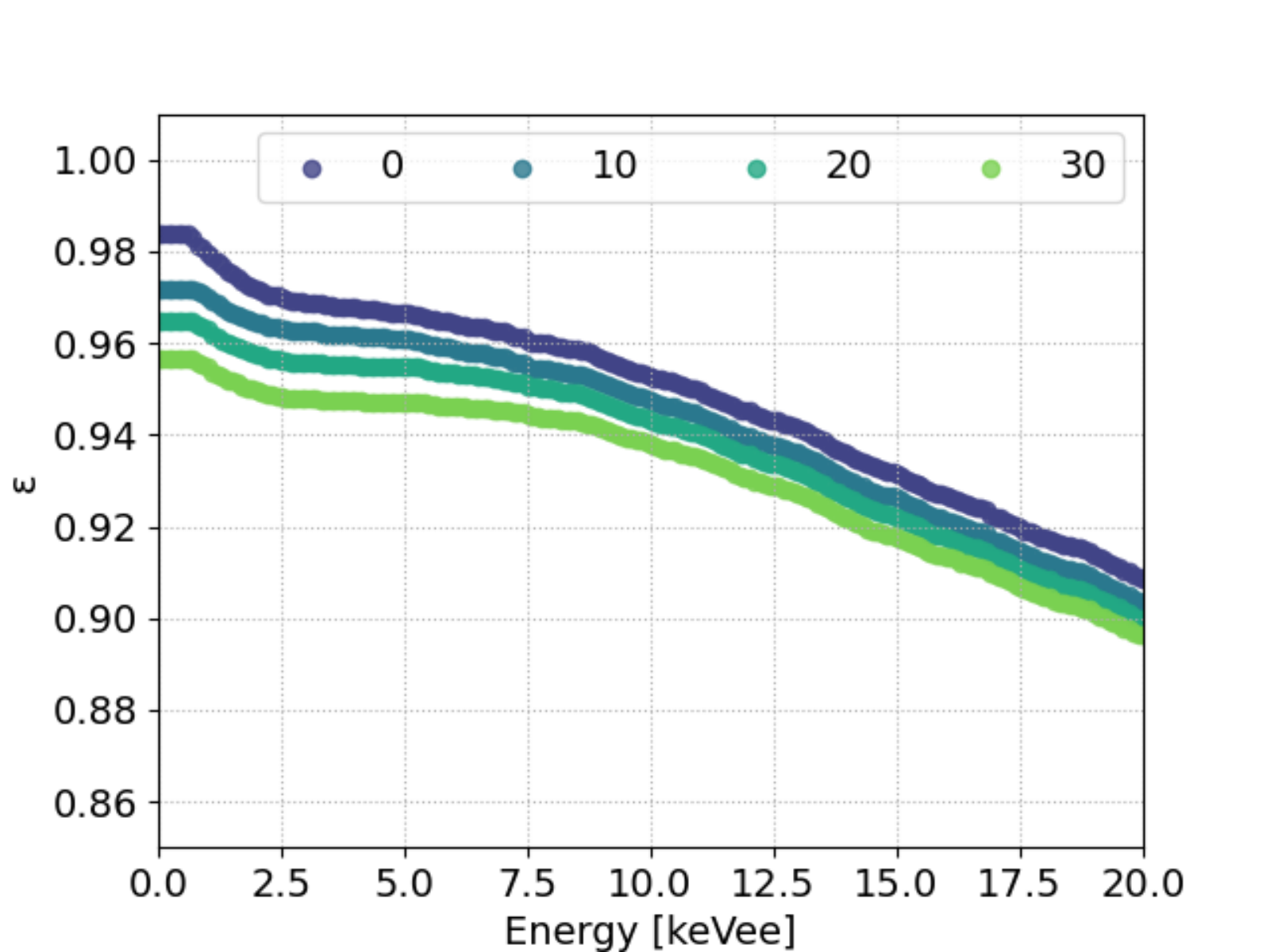}
\includegraphics[width=\columnwidth]{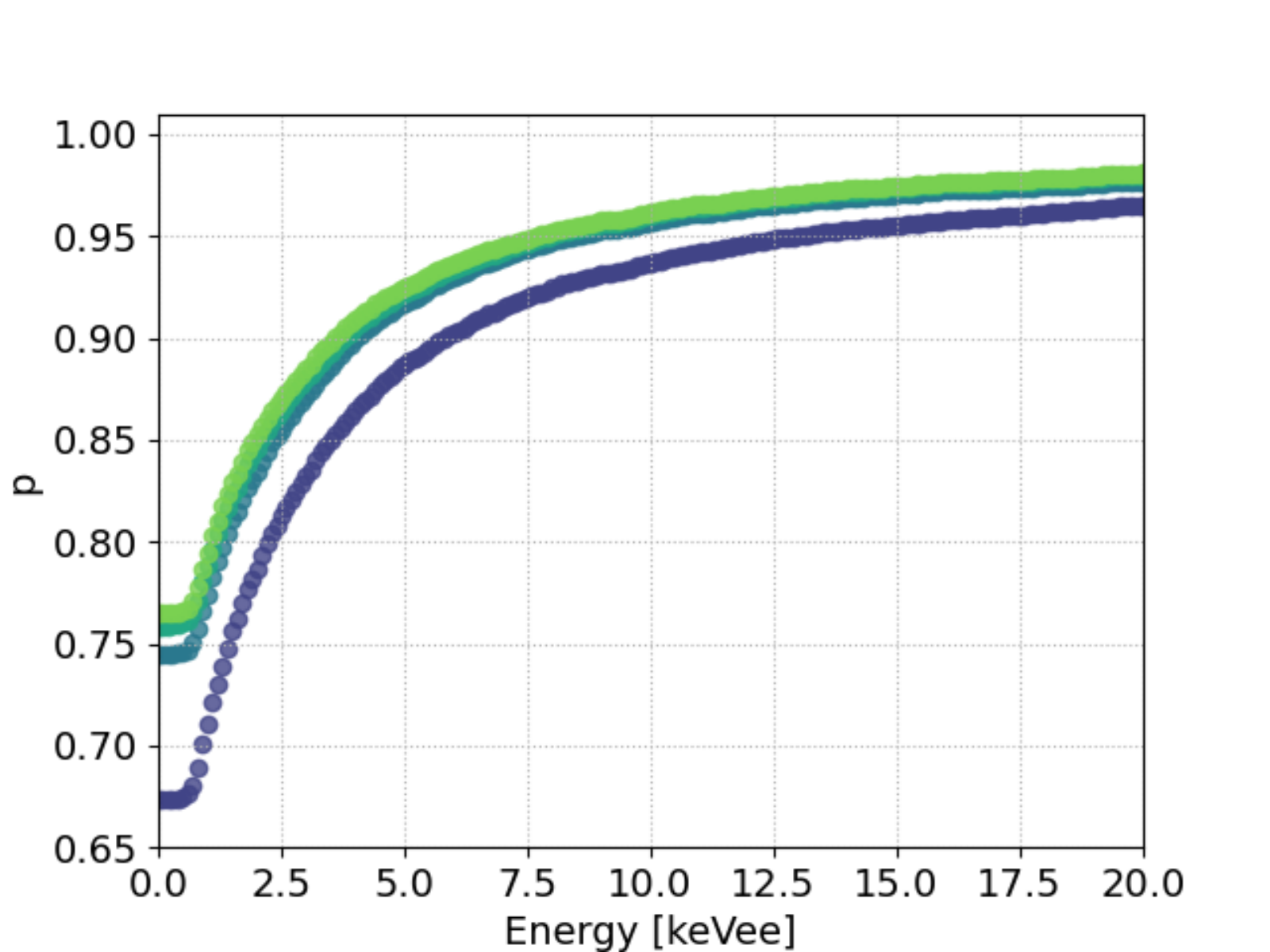}
\caption{Signal purity versus efficiency (top), efficiency versus reconstructed energy (middle), and purity versus reconstructed
energy (bottom) for different selections on $\mathrm{d}E/\mathrm{d}x$ to discriminate nuclear recoils from background. The colors indicate the lower threshold of the $\mathrm{d}E/\mathrm{d}x$ selection in units of keV$_{\rm ee}$/mm.}
\label{fig:lower-eff-pur-dEdx}
\end{figure}

For a single value of \(\mathrm{d}E/\mathrm{d}x\), we calculate the efficiency and purity above a minimum energy and define \(N_{\mathrm{sig}}\) to be the sum of helium, carbon, and oxygen recoils. The minimum energy values range from 1–100 keV$_{\rm ee}$. This process traces out a curve in \(p\) versus \(\epsilon\). This is shown in Fig. \ref{fig:lower-eff-pur-dEdx}. In this plot, a perfect selection would correspond to \(\epsilon = p = 1\). We thus define the optimum as the curve in $p$ versus $\epsilon$ with the smallest distance to $(1, 1)$. We find that the optimal selection corresponds to \(\mathrm{d}E/\mathrm{d}x > 10\) keV$_{\rm ee}$/mm with an energy threshold of 10.5 keV$_{\rm ee}$. Additionally, \(\epsilon\) and \(p\) are shown separately versus energy. As expected, \(p\) improves and \(\epsilon\) worsens with increasing energy and increasing \(\mathrm{d}E/\mathrm{d}x\). Using these selections, we find that we achieve \(\epsilon\) \(\sim\)0.95 and \(p\) \(\sim\)0.96 for nuclear recoils with electron and proton backgrounds.

\begin{figure}
\includegraphics[width=\columnwidth]{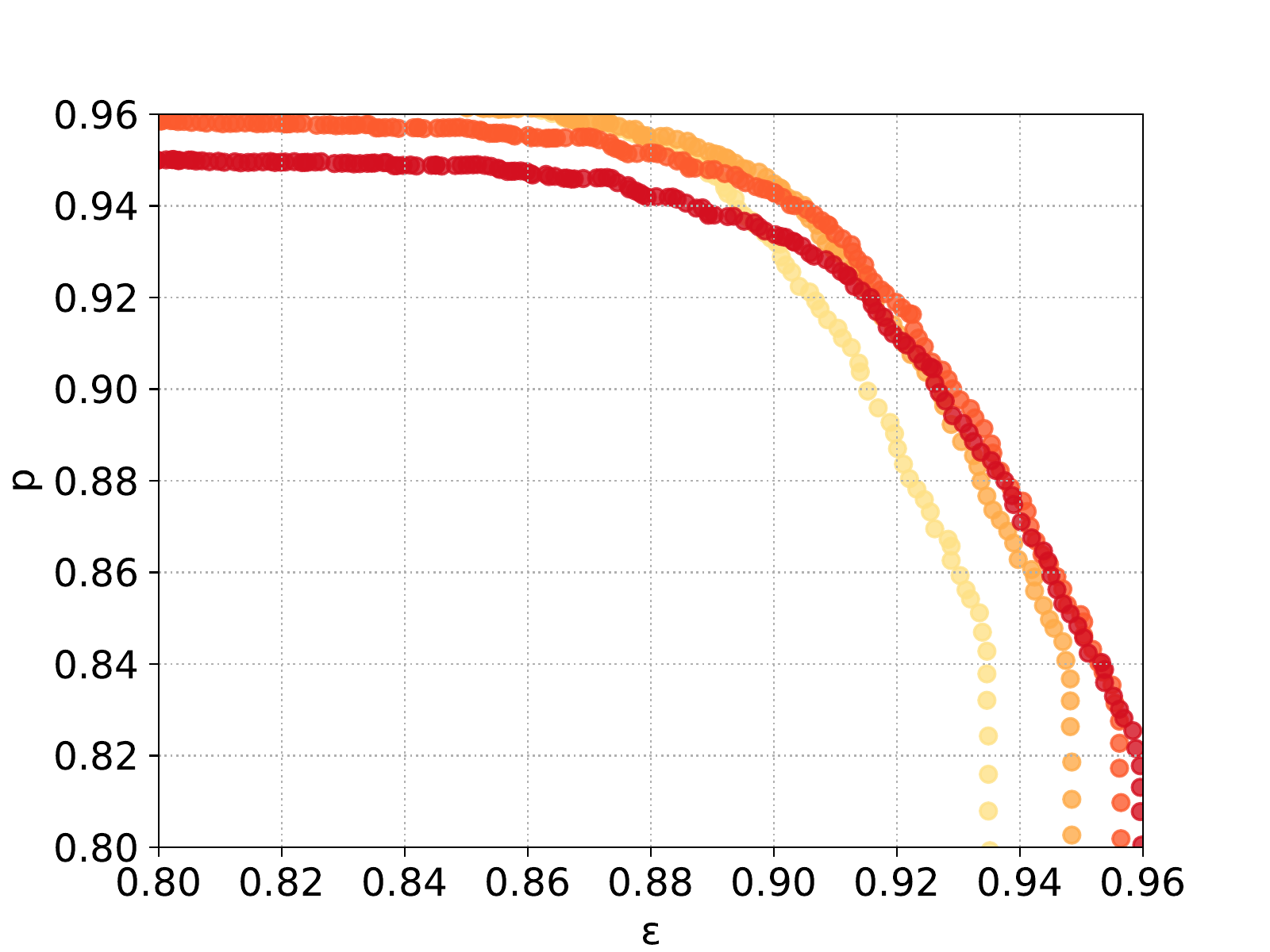}
\includegraphics[width=\columnwidth]{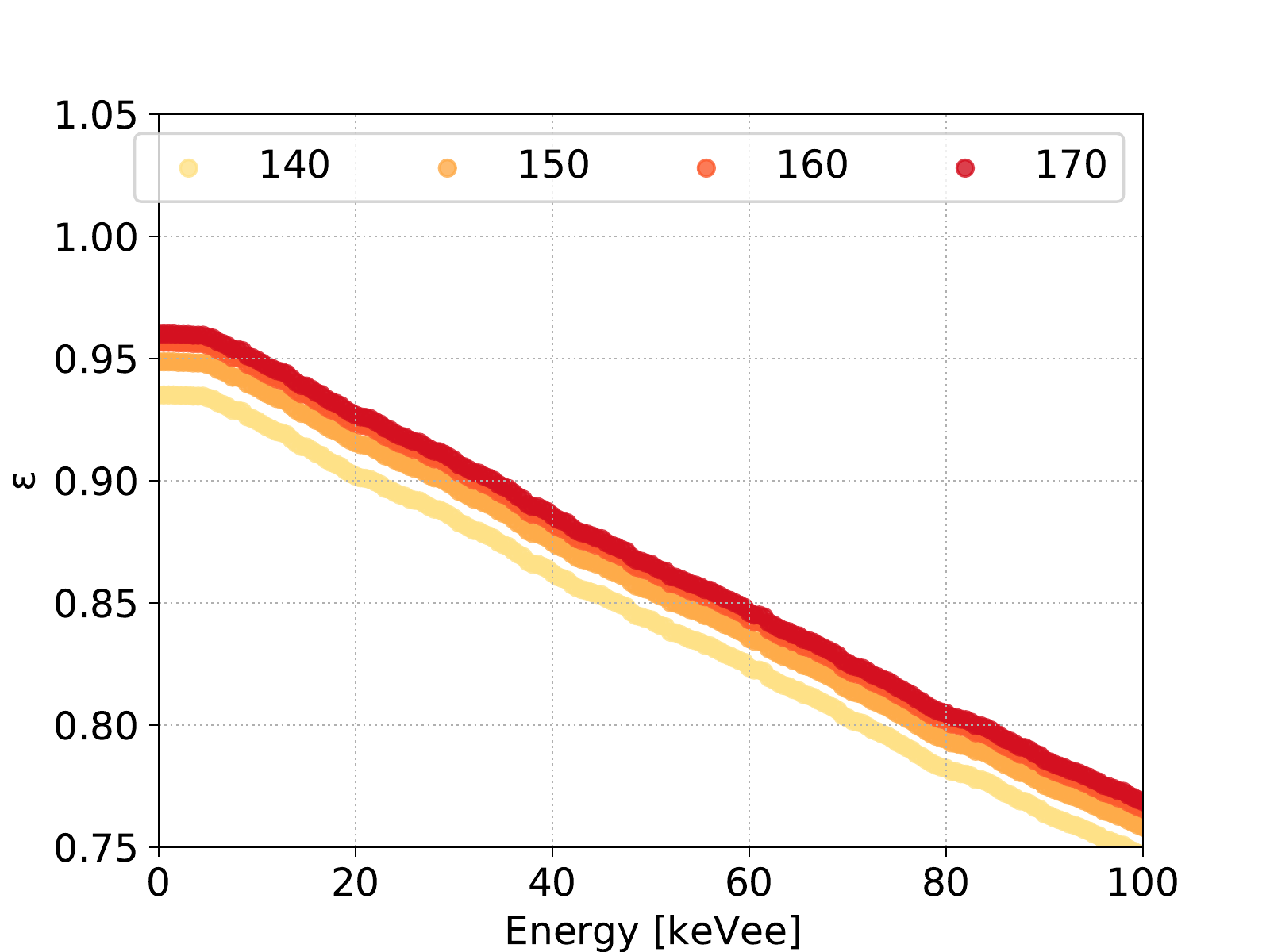}
\includegraphics[width=\columnwidth]{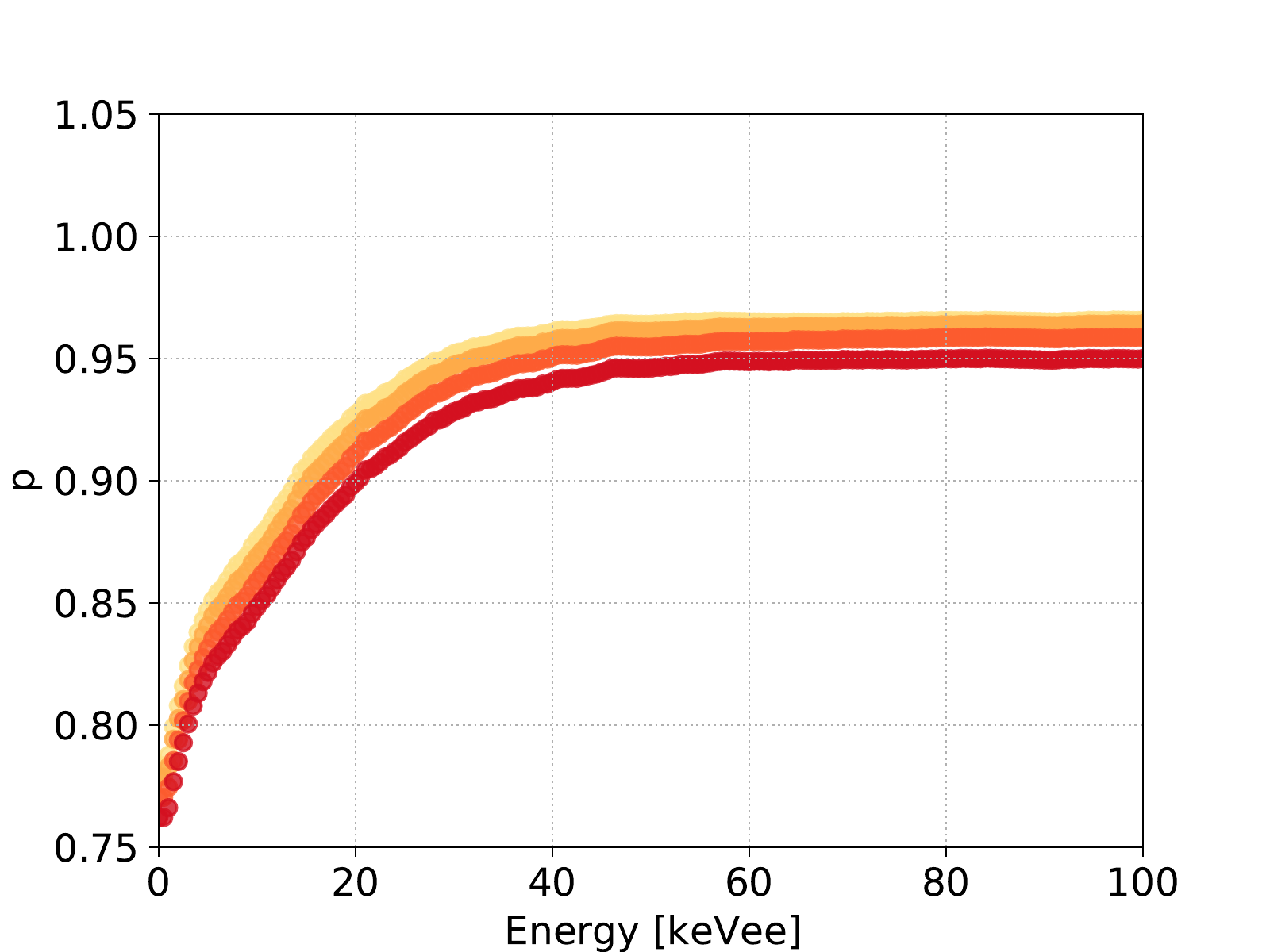}
\caption{Signal purity versus efficiency (top), efficiency versus reconstructed energy threshold (middle), and purity versus reconstructed
energy threshold (bottom) for different selections in $\mathrm{d}E/\mathrm{d}x$ to discriminate helium recoils from carbon and oxygen recoils. The colors indicate the upper threshold of the $\mathrm{d}E/\mathrm{d}x$ selection.}
\label{fig:upper-eff-pur-dEdx}
\end{figure}

This same procedure can be used to select an upper \(\mathrm{d}E/\mathrm{d}x\) to discriminate carbon and oxygen recoils from helium recoils. This is performed similarly to finding the lower \(\mathrm{d}E/\mathrm{d}x\) bound, with the exception that each point corresponds to a selection above a minimum and below a maximum value of \(\mathrm{d}E/\mathrm{d}x\). Furthermore, for this analysis, \(N_{\mathrm{sig}}\) corresponds to helium recoils only. From the previous analysis in MC, we choose the lower bound to be \(\mathrm{d}E/\mathrm{d}x > 10\) keV$_{\rm ee}$/mm, and we scan the upper limit using a range of minimum recoil energies from 1– 100 keV$_{\rm ee}$, as before. The efficiency versus purity, efficiency versus energy, and purity versus energy for four values of \(\mathrm{d}E/\mathrm{d}x\) are shown in Fig. \ref{fig:upper-eff-pur-dEdx}. We find that optimal efficiency and purity is reached by selecting events below 160 keV$_{\rm ee}$/mm above an energy threshold of 26.5 keV$_{\rm ee}$.

\begin{figure}
\includegraphics[width=\columnwidth]{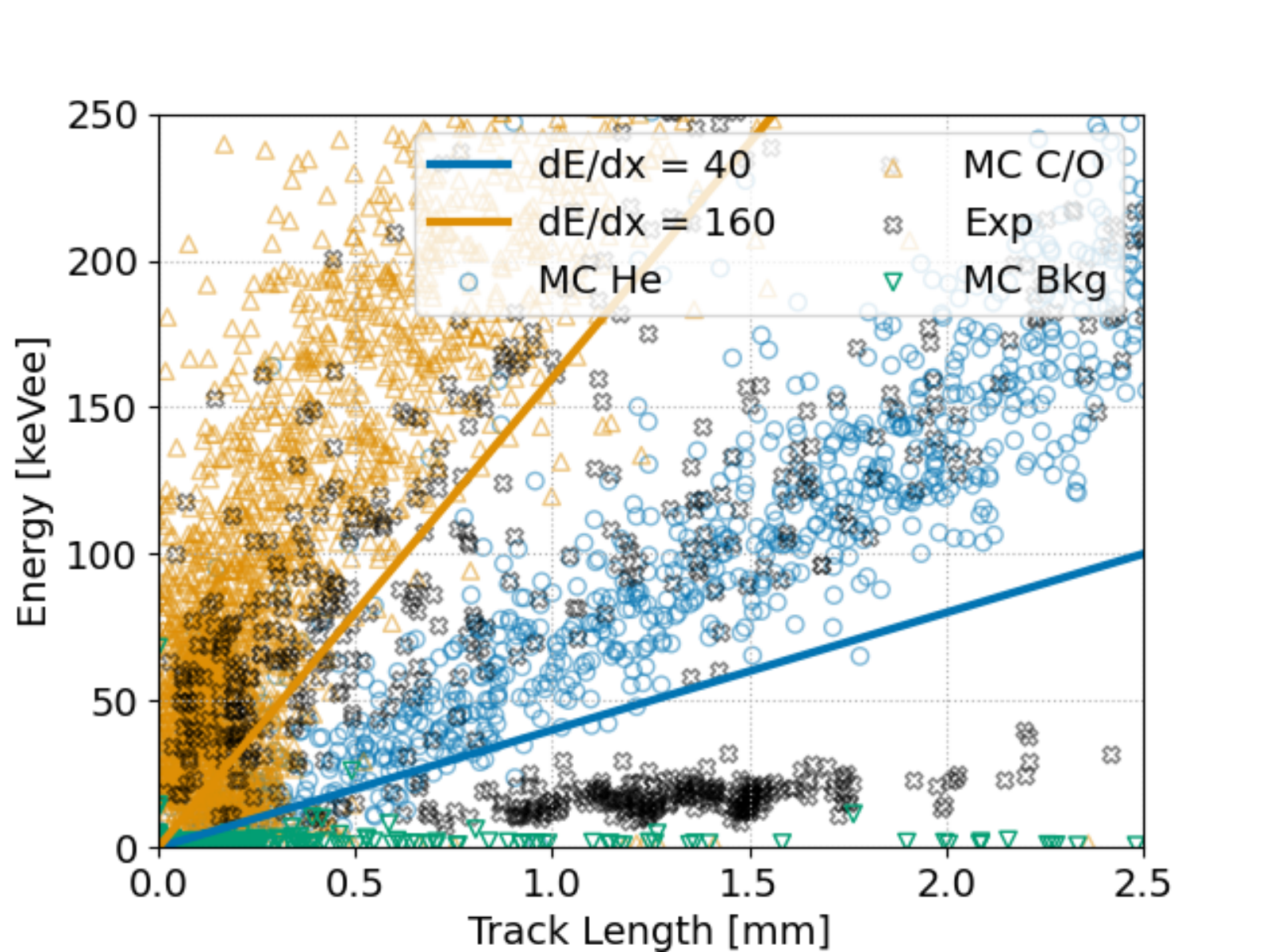}
\includegraphics[width=\columnwidth]{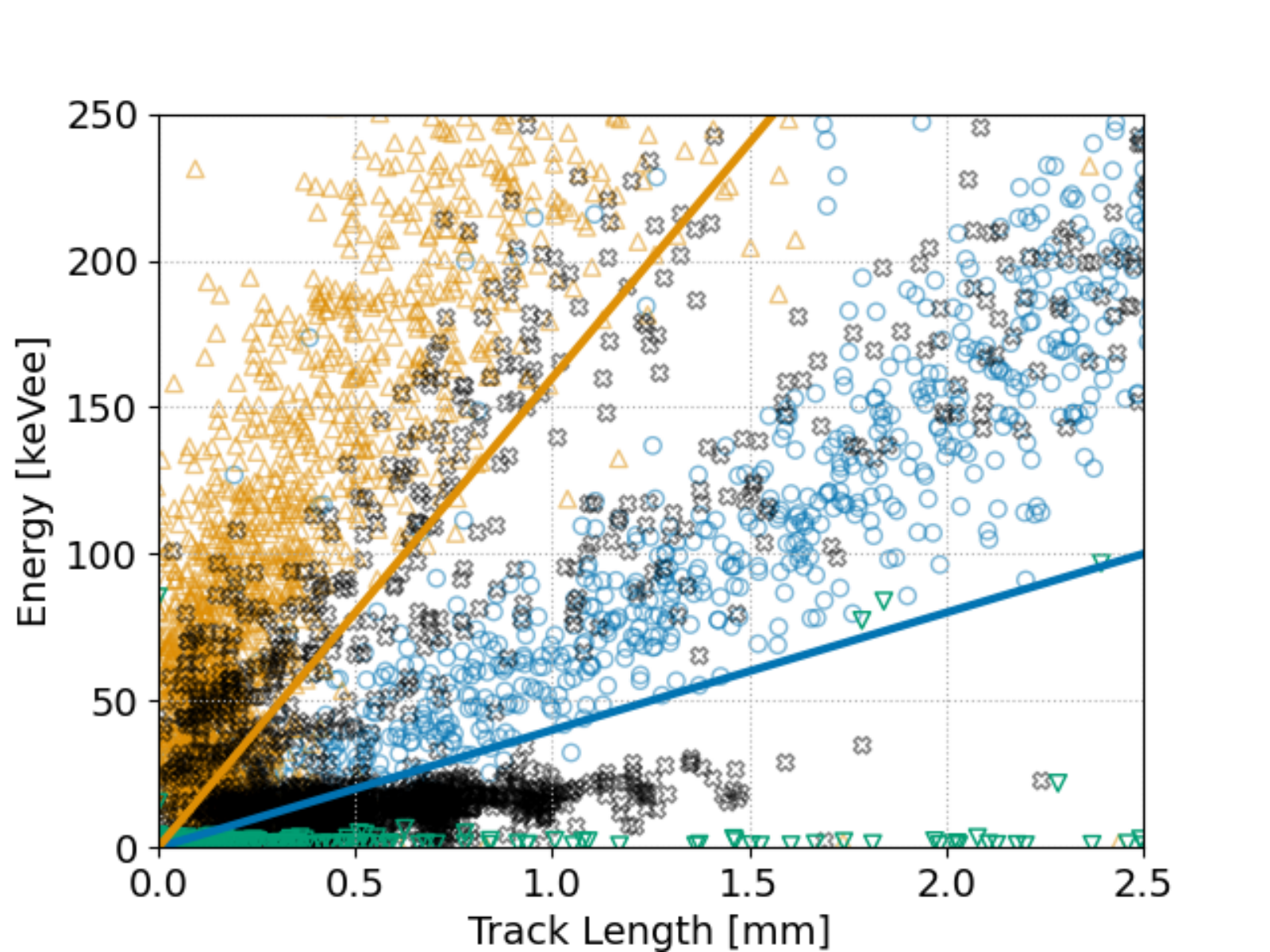}
\caption{Energy versus length for all events that pass the fiducialization selection in TPC H (top) and TPC V (bottom).
The blue line shows the boundary of a selection of $\mathrm{d}E/\mathrm{d}x > 40$ keV$_{\rm ee}$/mm, and the orange line shows the boundary of a selection of $\mathrm{d}E/\mathrm{d}x > 160$~keV$_{\rm ee}$/mm.}
\label{fig:tpc-EvsL}
\end{figure}

We can validate these selections in \(\mathrm{d}E/\mathrm{d}x\) by again looking at energy versus length. We begin by investigating the lower \(\mathrm{d}E/\mathrm{d}x\) selection used for discriminating low energy helium recoils from background events. As the background and helium bands are clearly visible, this check should determine if the \(\mathrm{d}E/\mathrm{d}x\) selection and/or an additional energy threshold should be applied to improve background rejection. We find that a selection of \(\mathrm{d}E/\mathrm{d}x > 40\) keV$_{\rm ee}$/mm with an energy threshold of 50 keV$_{\rm ee}$ provides a cleaner background rejection criteria. This is shown in Fig. \ref{fig:tpc-EvsL}. We note that there are significantly more events with \(\mathrm{d}E/\mathrm{d}x < 40 \)~keV$_{\rm ee}$/mm in TPC V. We have found that this was due to either a sub-optimal performance or a mistaken setting of the firmware-level background veto. However, as we see in Fig. \ref{fig:tpc-EvsL}, events in this region can be rejected without significant reduction in selecting the nuclear recoil signal. We also show the upper selection of \(\mathrm{d}E/\mathrm{d}x = 160 \)~keV\(_{\rm ee}\)/mm in Fig. \ref{fig:tpc-EvsL} and find that this value is sufficient for distinguishing helium recoils from carbon and oxygen recoils.  

\begin{figure}[htbp]
\centering
\includegraphics[width=.9\linewidth]{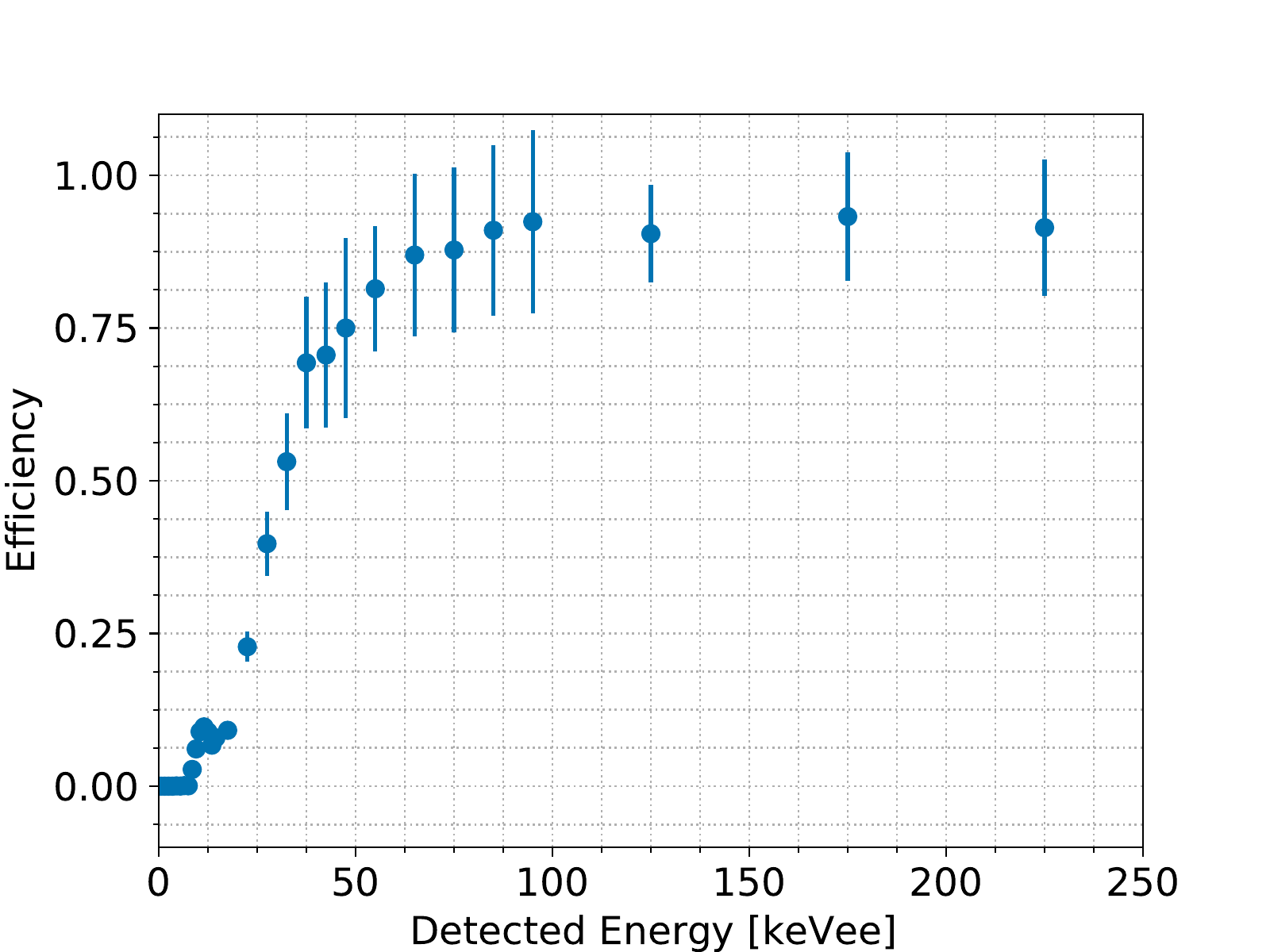}
\caption{\label{fig:sel-eff}Efficiency of TPC nuclear recoil selections versus detected energy in experimental data.}
\end{figure}

Using the obtained selection values for background rejection, we calculate the efficiency of the event selections in experimental data as the fraction of events passing the edge veto and the minimum \(\mathrm{d}E/\mathrm{d}x\) selection. This efficiency versus energy serves to determine which nuclear recoil energies we are sensitive to and whether the selections bias the observed energy spectrum. This efficiency is shown in Fig. \ref{fig:sel-eff}. The efficiency becomes 50\% at approximately 30 keV$_{\rm ee}$ and is near unity and flat for recoil energies larger than \(\sim\)65 keV$_{\rm ee}$.

\subsection{Directional performance: axial directionality}
\label{sec:orgb2efae9}

We begin by first measuring the accuracy of the reconstructed track axis, or \emph{axial directionality}, of events.
The initial track fit detects the recoil track axis, but is insensitive to the sign of the recoil track vector. We call this axial directionality. We evaluate performance at this level as follows. We choose the sign of the recoil track vector such that \(-90^{\circ} < \phi < 90^{\circ}\). We likewise obtain the true track axis by folding the Monte Carlo truth direction, and then calculate the 3D angle between the reconstructed track axis to the true track axis. Given that this is a test of axial directionality, the angle between the two axes cannot exceed \(90^{\circ}\). Performing this measurement on many tracks serves as a measure of the average of the true axial mismeasurement, or angular resolution, of the TPCs for a given track length.

\begin{figure}
\includegraphics[width=\columnwidth]{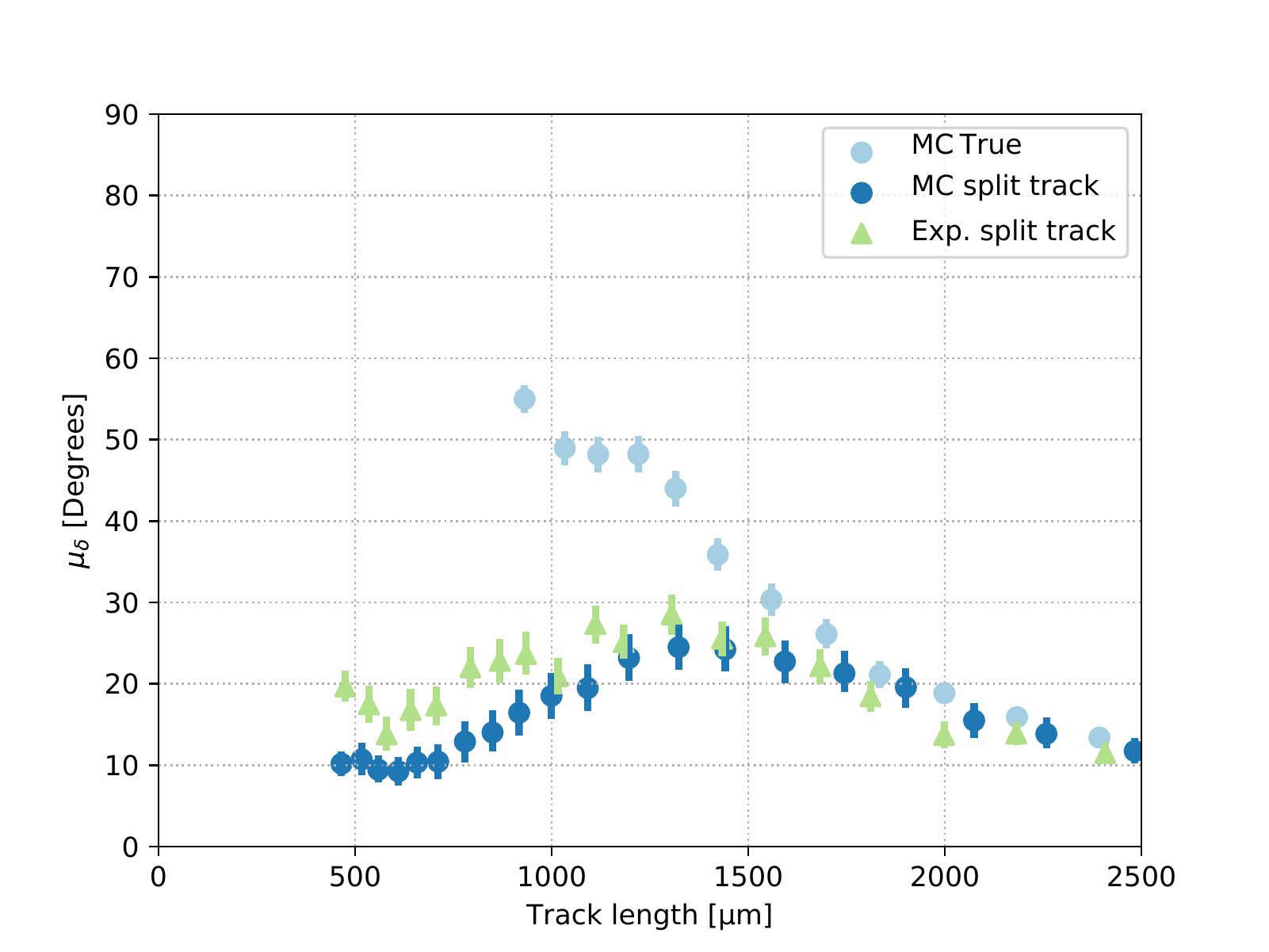}
\includegraphics[width=\columnwidth]{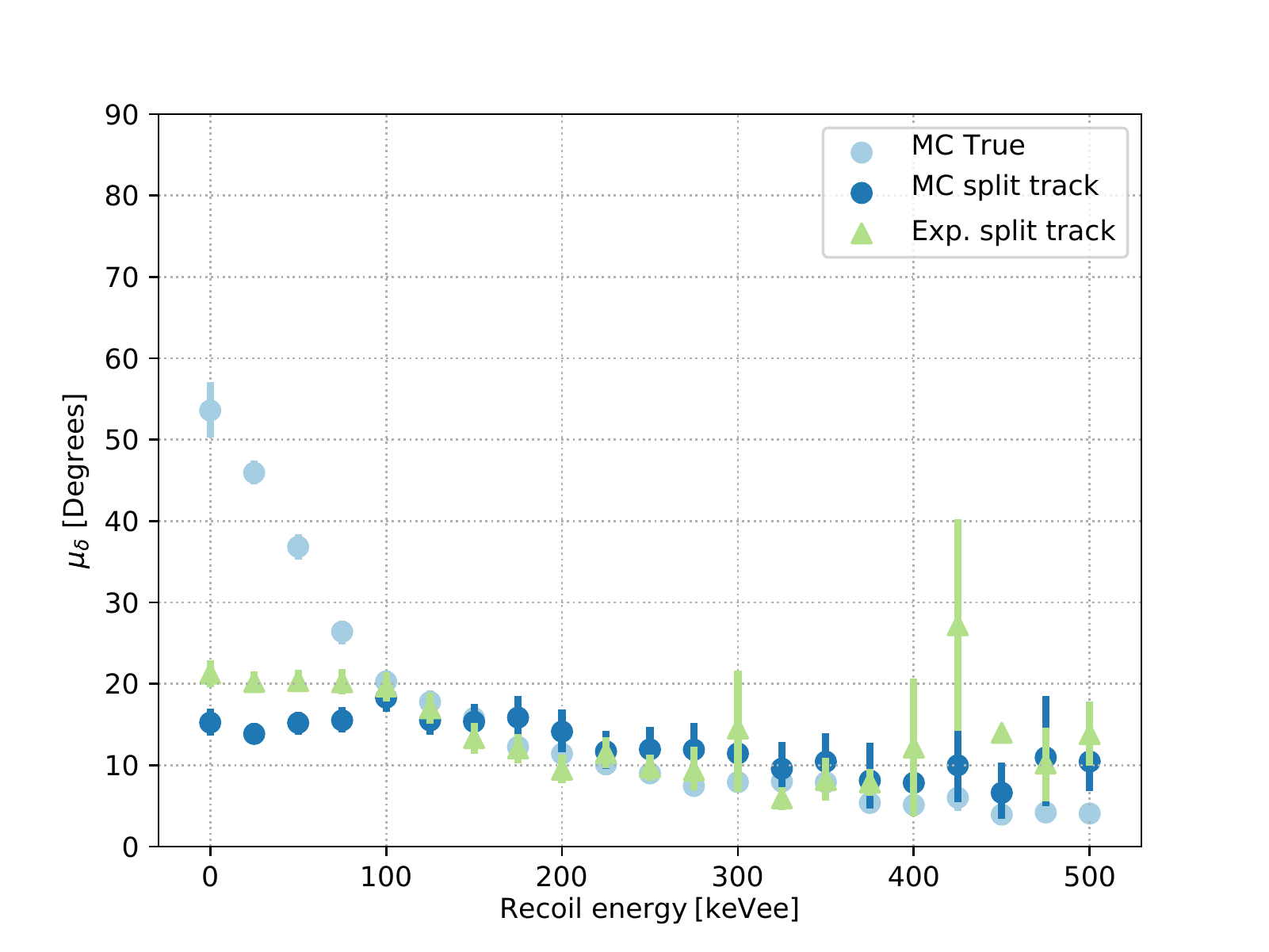}
\caption{Angular mismeasurement versus track length (top) and recoil energy (bottom) for the true mismeasurement of the full track (light blue circles) and the
split track mismeasurement in Monte Carlo data (dark blue circles) and experimental data (green triangles).}
\label{fig:angularRes_vs_Length}
\end{figure}

Ideally, we would like to have a method to determine the angular resolution that only requires experimental data. We attempt to do so by dividing each track into halves, bisecting the track at the midpoint along the reconstructed track axis. We then reconstruct a track axis for each half independently and calculate the 3D angle difference between the two halves and divide by \(\sqrt{2}\) to account for the propagated error associated with two fits. We then plot this quantity versus the length of the halved, or \emph{split}, track. The top plot of Fig. \ref{fig:angularRes_vs_Length} shows the mean angular difference per bin, \(\mu_{\delta}\), versus the length of the tracks used in each method. 
For the true mismeasurement (light blue circles), \(\mu_{\delta}\) is calculated using the full reconstructed axial track and the angle of the true track vector from Monte Carlo. The dark blue circles and the green triangles represent the intra-track, or split track, mismeasurement versus the length of the split track. The bottom plot shows \(\mu_{\delta}\) versus reconstructed energy of the event. In all cases, the error bars represent one standard deviation of the distribution in each bin in the horizontal axis. From these figures, we find that the intra-track mismeasurement obtained from halving a single track does agree with the true mismeasurement in both Monte Carlo and experimental data at energies above 100 keV$_{\rm ee}$with an angular resolution of \(20^{\circ}\).

\subsection{Directional performance: vector directionality}
Next we seek to infer the \emph{vector direction} of recoils. We take each obtained track axis and assign a vector direction to that axis by utilizing the specific energy loss of the recoil. In general, the specific energy loss can change significantly along the recoil trajectory depending on the initial recoil energy and the distance it travels within a medium. However, we only consider events that pass the fiducialization criteria mentioned in Section \ref{sec:sels}, which requires the entire path of the recoil to be contained within the fiducial volume of the TPC. This means that we only select events such that specific energy loss of the recoil will lie at the end of the Bragg curve, corresponding to a sharp decline in the specific energy loss until the recoil stops. Therefore, in principle, one should be able to identify the positive vector trajectory of the track, often referred to as the \emph{head}, as the end with less detected energy, and vice-versa for the \emph{tail} of the track.

\begin{figure}
\centering
\includegraphics[width=\columnwidth]{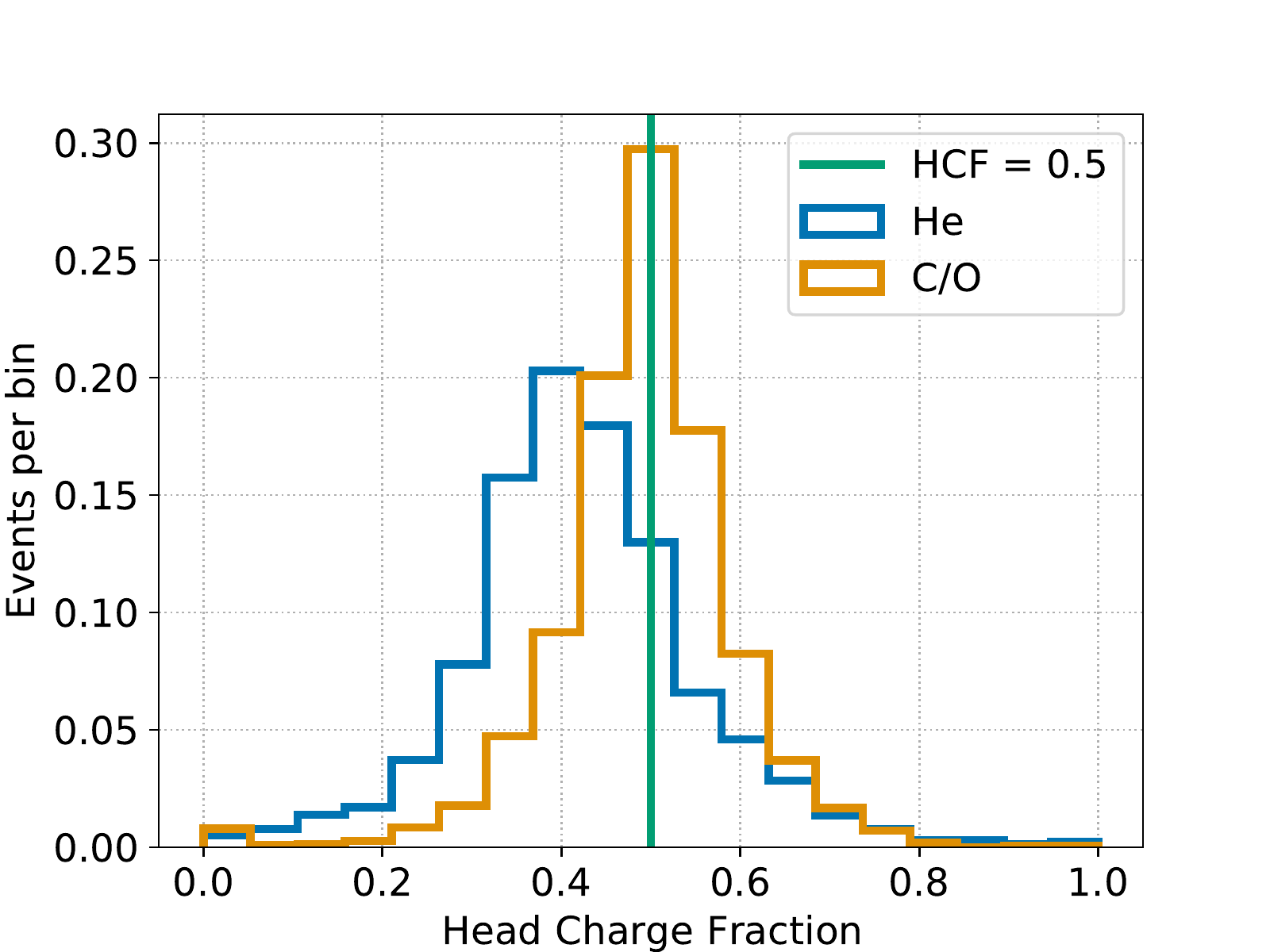}
\caption{Fractional charge in the true head of helium events in blue and in carbon and oxygen events in orange.}
\label{fig:true_hcf_tpcH}
\end{figure}

To test this, we take the simple approach of dividing the track into two halves and use the truth information in the Monte Carlo data to plot the ratio of detected energy in the true head to the total amount of detected charge in an event. This is shown in Fig. \ref{fig:true_hcf_tpcH} for helium events in blue and for carbon and oxygen events in orange. We note in these plots that the helium recoils exhibit the expected behavior—less than half of the detected charge, or energy, is found in the head, corresponding to a head charge fraction (HCF) of less than 0.5. For carbon and oxygen, however, the HCF distribution peaks at HCF = 0.5, meaning that the head of the track is, on average, indistinguishable from the tail of the track. Thus, we conclude that vector directionality using this method is only effective for helium recoils. Using a selection of HCF < 0.5, we obtain a head-tail selection efficiency of 0.78 for events for helium recoils.

\begin{figure}
\centering
\includegraphics[width=\columnwidth]{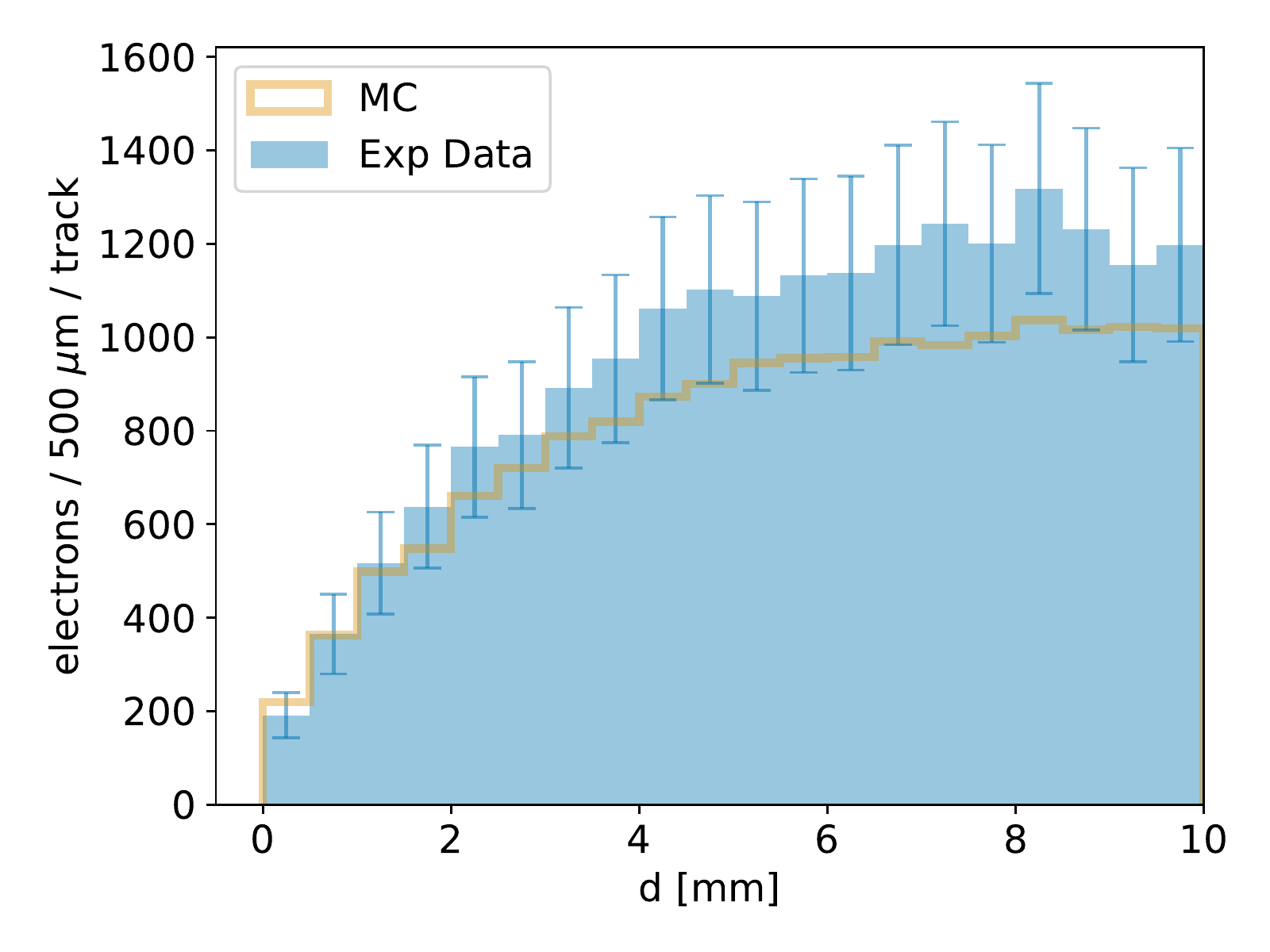}
\caption{Detected charge versus distance from the track head in selected helium recoils. The orange line corresponds to a digitization of SRIM-based events in simulated data and the blue histogram corresponds to the equivalent measurement in data. The error bars show the statistical variation in the charge density in the experimental track sample analyzed.}
\label{fig:distFromHead}
\end{figure}

To further verify vector directionality, we can check whether the amount of detected charge along the recoil path agrees in simulation and experiment. We do so by selecting long helium recoil events and define the head of the track as the half with less charge. We then make a histogram of the gain-corrected detected charge versus the distance from the track head.  These histograms are shown in Fig. \ref{fig:distFromHead}.

The simulation, shown as an orange line in the figure, uses SRIM \cite{SRIM} to model the energy loss of helium recoils in the He:CO$_2$ target gas. The blue histogram shows the observed distribution in experimental data. We find that the longitudinal charge distribution of tracks predicted by SRIM agrees rather well, though not perfectly, with the measured asymmetry.

We note that this comparison is sensitive to both the ionization distributions along the track and to the detector gain, which affect the shape and normalization of the curves, respectively. The blue distribution includes the fixed 20\% gain correction factor for experimental data outlined in Sec. \ref{sec:sels}. This factor improves agreement in simulated and experimental data for the majority of recorded events, which are of lower energy and lengths than those shown in Fig. \ref{fig:distFromHead}. We speculate that having normalized our gain based on shorter tracks could be a contributing factor to the disagreement between simulated and experimental data at larger lengths. If we instead float the normalization in our comparison, and only compare simulated and observed ionization distribution shapes, the agreement improves further.

\begin{figure*}
\centering
\begin{subfigure}{0.5\textwidth}
  \centering
  \includegraphics[width=\textwidth]{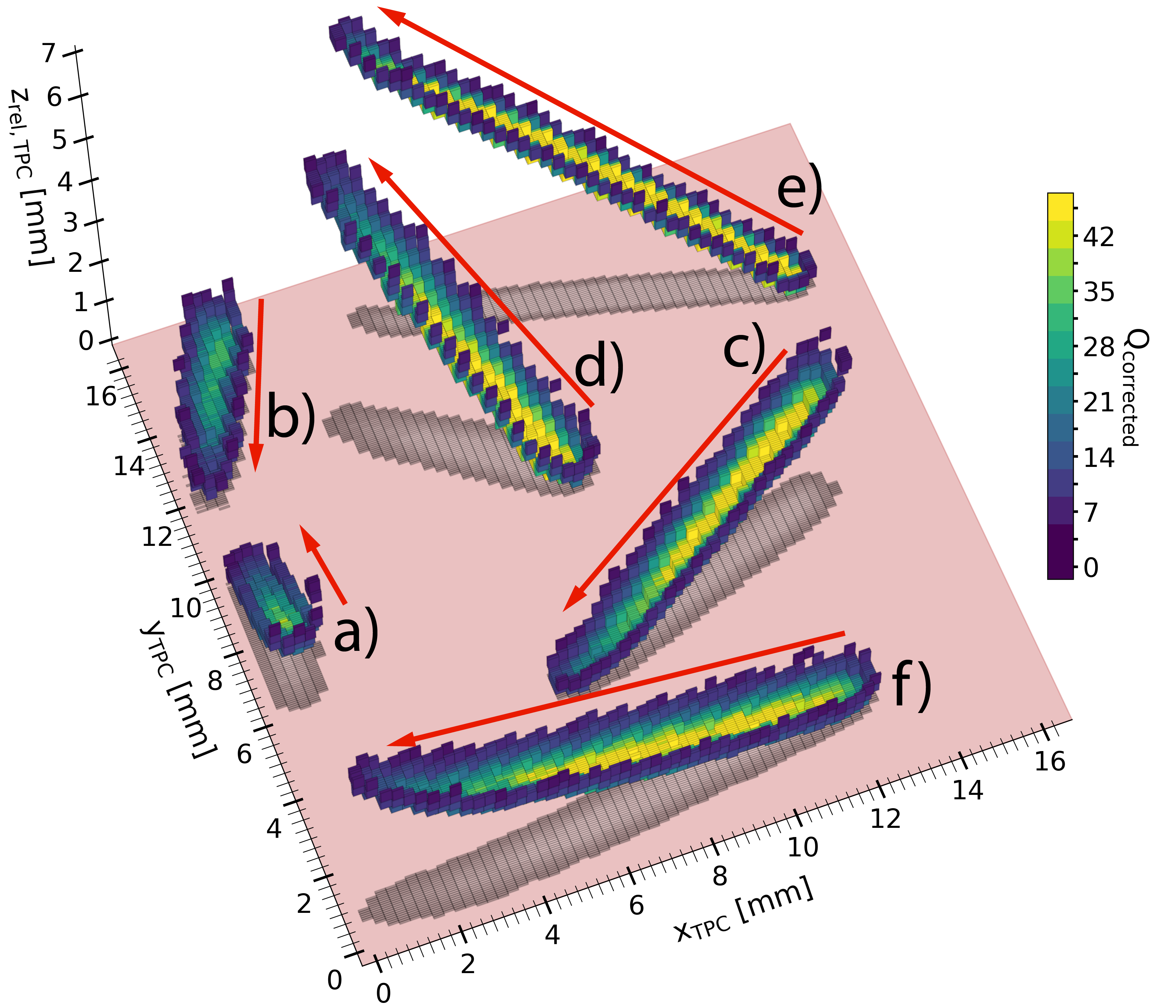}
  \caption{}
\end{subfigure}
\begin{subfigure}{0.45\textwidth}
  \centering
  \includegraphics[width=0.9\textwidth]{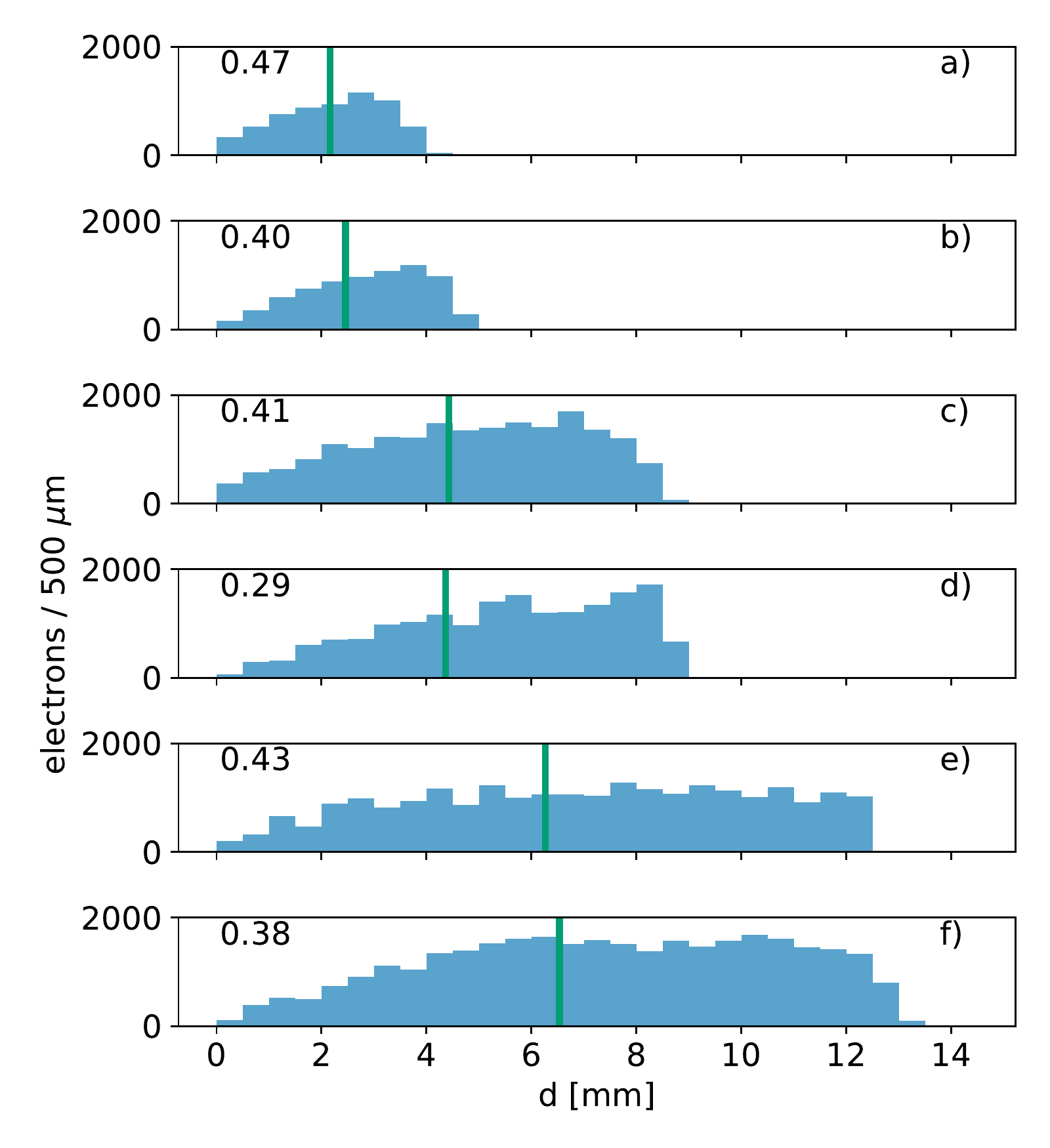}
  \caption{}
\end{subfigure}
\caption{Six tracks visualized in 3D (a) alongside their charge distributions versus distance from the track head (b). The head-direction of the tracks is shown with red arrows, and is determined by designating the half with less charge as the head, as shown by the color scale. On the right, the geometric midpoint of each track is shown as a vertical green bar. The number in the upper left displays the head charge-fraction of the track.}
\label{fig:display-distFromHead}
\end{figure*}

While the charge asymmetry is clear when many tracks are averaged, it is also  interesting to view individual tracks from experimental data, to gauge the typical event-level strength of the charge asymmetry. Fig. \ref{fig:display-distFromHead} shows six tracks, labelled a) -- f), plotted in a 3D event display on the left alongside their longitudinal charge distribution.

The red arrows on the left plot point in the direction of the head, which corresponds to the end of the track with less charge, as indicated by the color. On the right, the distribution of charge versus $d$ is shown in the blue histogram. The green bars correspond to the half-length of the track, and the number in the top left corresponds to the HCF, which is the amount of charge to the left of the green line divided by the sum of the total charge in the event.

In Figs. \ref{fig:distFromHead} and \ref{fig:display-distFromHead}, we see a consistent behavior in both simulated and experimental data---the charge distribution gradually decreases to zero as the distance from the head decreases. The signature is clear even at the individual event-level.

\section{Analysis of SuperKEKB Phase 1 fast neutron events}
\label{sec:phase1-ana}
The updated performance analysis motivates an extension of the analysis presented in Ref.~\cite{lewis19_first_measur_beam_backg_at_super}. As such, we discuss simulation and event selections, experimental, runs for the TPCs, energy spectra of nuclear recoils, nuclear recoil event rate versus beam size, and directional analysis of nuclear recoils.

\subsection{Simulation and event selections}
\label{sec:org45d4f85}
The details of the procedure of producing the BEAST II Monte Carlo data are described in Section 4 of Ref \cite{lewis19_first_measur_beam_backg_at_super}. The general production pipeline consists of: (a) generation of beam particles from SuperKEKB beam-induced backgrounds, (b) modeling of the transport and interactions of primary and secondary particles, (c) simulating the detector response and digitization, and (d) scaling of the detector response with accelerator conditions present during experimental runs.

The Strategic Accelerator Design (SAD) software framework \cite{SADHP} performs step (a) and Geant4 performs steps (b) through (c). The SAD simulation is performed such that the beam loss rates correspond to one second of simulated beam time for the SuperKEKB accelerator with constant beam parameters listed in Table \ref{tbl:MC-beam-params}. The primary particles are then passed through steps (b) and (c). The simulated TPC data contains digitized events that replicate the data format produced by the FE-I4B chip output. This series of steps translates to a simulated rate in the TPCs that can then be scaled with beam parameters.

Due to the small interaction probability of neutrons in the TPCs, it is necessary to provide longer simulation times in order to acquire a sufficiently large simulated data sample. This is done by simulating each neutron that enters the TPC volume 18,000 times—resulting in a 5 hours-equivalent sample. This results in a total of 13,011 simulated events for fast neutron analysis at the default beam parameters in Table \ref{tbl:MC-beam-params}.

The event rate in the simulated data must then be reweighted to the beam conditions measured during the experimental data runs to achieve step (d). As shown in Ref.\cite{lewis19_first_measur_beam_backg_at_super}, the Touschek background scales with \(I^2/\sigma_y\)  and the beam-gas background scales with \(IPZ_e^2\), where \(I,~\sigma_y,~P,\) and \(Z_e\) are the beam current, beam size in the \(y\)-direction as measured by the X-ray monitors \cite{Flanagan:2005}, average pressure of the residual gas in the ring, and effective atomic number of the residual gas in the ring, respectively.

\begin{table}[htbp]
\caption{\label{tbl:MC-beam-params}Beam parameters used for SuperKEKB Phase 1 Monte Carlo simulated data \cite{lewis19_first_measur_beam_backg_at_super}.}
\centering
\begin{tabular}{rrr}
Machine Parameters & HER & LER\\
\hline
Beam current \(I\) [A] & 1.0 & 1.0\\
Number of bunches \(N_b\) & 1000 & 1000\\
Bunch current \(I_b\) [mA] & 1.0 & 1.0\\
Vertical beam size \(\sigma_y\) [\si\micro m] & 59 & 110\\
Emittance ratio \(\epsilon_y/\epsilon_x\) & 0.1 & 0.1\\
Pressure \(P\) [nTorr] & 10 & 10\\
\end{tabular}
\end{table}

All TPC event selections and their cumulative efficiencies on simulated and experimental data for all analyses are
shown in Table \ref{tbl:all-sel-effs}. Each analysis will use at least the first three selections, which are necessary for nuclear recoil
selection and background rejection. The remaining selections are only required to isolate helium recoils for directional analyses.

\begin{table}[htbp]
\caption{\label{tbl:all-sel-effs}Event selections used in all analyses, along with each selection's cumulative efficiency in MC Touschek (T), MC beam-gas (BG), and experimental data in TPCs H and V separately.}
\centering
\begin{tabular}{rrrr}
TPC H & MC T & MC BG & Exp\\
\hline
Edge veto & 0.46 & 0.48 & 0.0444\\
\(\mathrm{d}E/\mathrm{d}x > 40\) keV$_{\rm ee}$/mm & 0.31 & 0.32 & 0.0426\\
\(E > 50\) keV$_{\rm ee}$& 0.23 & 0.25 & 0.0419\\
 &  &  & \\
\emph{He recoil selections} &  &  & \\
\(\mathrm{d}E/\mathrm{d}x < 160\) keV$_{\rm ee}$/mm & 0.12 & 0.13 & 0.0027\\
\(|\phi_{\rm axial}| > 160^{\circ}\) & 0.03 & 0.03 & 0.0007\\
HCF < 0.5 & 0.02 & 0.02 & 0.0005\\
 &  &  & \\
TPC V &  &  & \\
\hline
Edge veto & 0.48 & 0.46 & 0.1333\\
\(\mathrm{d}E/\mathrm{d}x > 40\) keV$_{\rm ee}$/mm & 0.30 & 0.31 & 0.1265\\
\(E > 50\) keV$_{\rm ee}$& 0.24 & 0.25 & 0.0052\\
 &  &  & \\
\emph{He recoil selections} &  &  & \\
\(\mathrm{d}E/\mathrm{d}x < 160\) keV$_{\rm ee}$/mm & 0.12 & 0.12 & 0.0030\\
\(-110^{\circ} < \phi_{\rm axial} < -70^{\circ}\) & 0.03 & 0.03 & 0.0008\\
HCF < 0.5 & 0.02 & 0.02 & 0.0005\\
\end{tabular}
\end{table}

\subsection{Experimental runs for TPCs}
\label{sec:orgb008e48}
For the fast neutron measurements in experimental data, we performed dedicated, longer-duration runs specifically to accumulate a sufficient sample of nuclear recoils in the TPCs. A run for the HER occurred on May 23, 2016 for approximately 1.5 hours at an average measured vertical beam size of approximately 40 \si\micro m with initial beam current of 500 mA. Table \ref{tbl:HER-run} shows the number of detected events that pass the first three selections in Table \ref{tbl:all-sel-effs} compared to the reweighted number of Monte Carlo events passing the same selections for this run.

We find that the Monte Carlo underestimates the observed number of events recorded by the TPCs by approximately a factor of five in both TPCs, with a very large uncertainty due to limited statistics. Studies in later phases of BEAST II have indicated that this discrepancy between predicted and observed rates in the HER is most likely due to improper collimator simulations in the HER. A more thorough analysis of this effect will be studied and presented in future results \cite{andrii}. For the scope of this result, we note the effect, but do not correct for it.

\begin{table}[htbp]
\caption{\label{tbl:HER-run}Number of events predicted by Monte Carlo compared to experimental measurement for the HER run.}
\centering
\begin{tabular}{rrr}
 & TPC H & TPC V\\
\hline
MC Beam-gas & 3 \textpm{} 0 & 4 \textpm{} 0\\
MC Touschek & 3 \textpm{} 1 & 4 \textpm{} 1\\
Experiment & 46 \textpm{} 7 & 36 \textpm{} 6\\
\end{tabular}
\end{table}

Due to the fact that the Touschek contribution to beam backgrounds is predicted to be far more problematic in the LER than in the HER and given the very low detection rate of the TPCs, it was decided to devote substantially more experiment time to collecting data from the LER than for the HER for fast neutron analysis. The resulting larger statistics allow us to perform more detailed studies of the neutron background from the LER, including studies of directional distributions and separating the beam-gas and Touschek contributions to the background in experimental data.

Dedicated LER runs occurred on May 29, 2016 for approximately 5.5 hours at a beam current of approximately 600 mA, topping off the beam as required. The beam size was set at three specific values by intentionally increasing the beam emittance and therefore the vertical beam size. Each run corresponds to one beam size setting. The beam size was measured using the X-ray monitors \cite{Flanagan:2005}, and was measured to be approximately 40 \si\micro m, 60 \si\micro m, and 90 \si\micro m for the three runs, respectively. Each run is further divided into sub-runs. A sub-run is defined as a period of time of stable beam conditions at the desired settings as defined above, specifically excluding injection times. Table \ref{tbl:LER-run} shows the number of detected events compared to the reweighted Monte Carlo prediction for this run. We find that for the LER the agreement between simulation and experimental data is better. On average, the observed number of events is consistent with the prediction for TPC V and $\sim40\%$ lower than the prediction in TPC H.

\begin{table}[htbp]
\caption{\label{tbl:LER-run}Number of events predicted by Monte Carlo compared to experimental measurement for the LER run.}
\centering
\begin{tabular}{rrr}
 & TPC H & TPC V\\
\hline
MC Beam-gas & 340 \textpm{} 7 & 272 \textpm{} 6\\
MC Touschek & 536 \textpm{} 16 & 412 \textpm{} 10\\
Experiment & 567 \textpm{} 22 & 640 \textpm{} 80\\
\end{tabular}
\end{table}

\subsection{Energy spectra of nuclear recoils}
\begin{figure}
\centering
\includegraphics[width=\columnwidth]{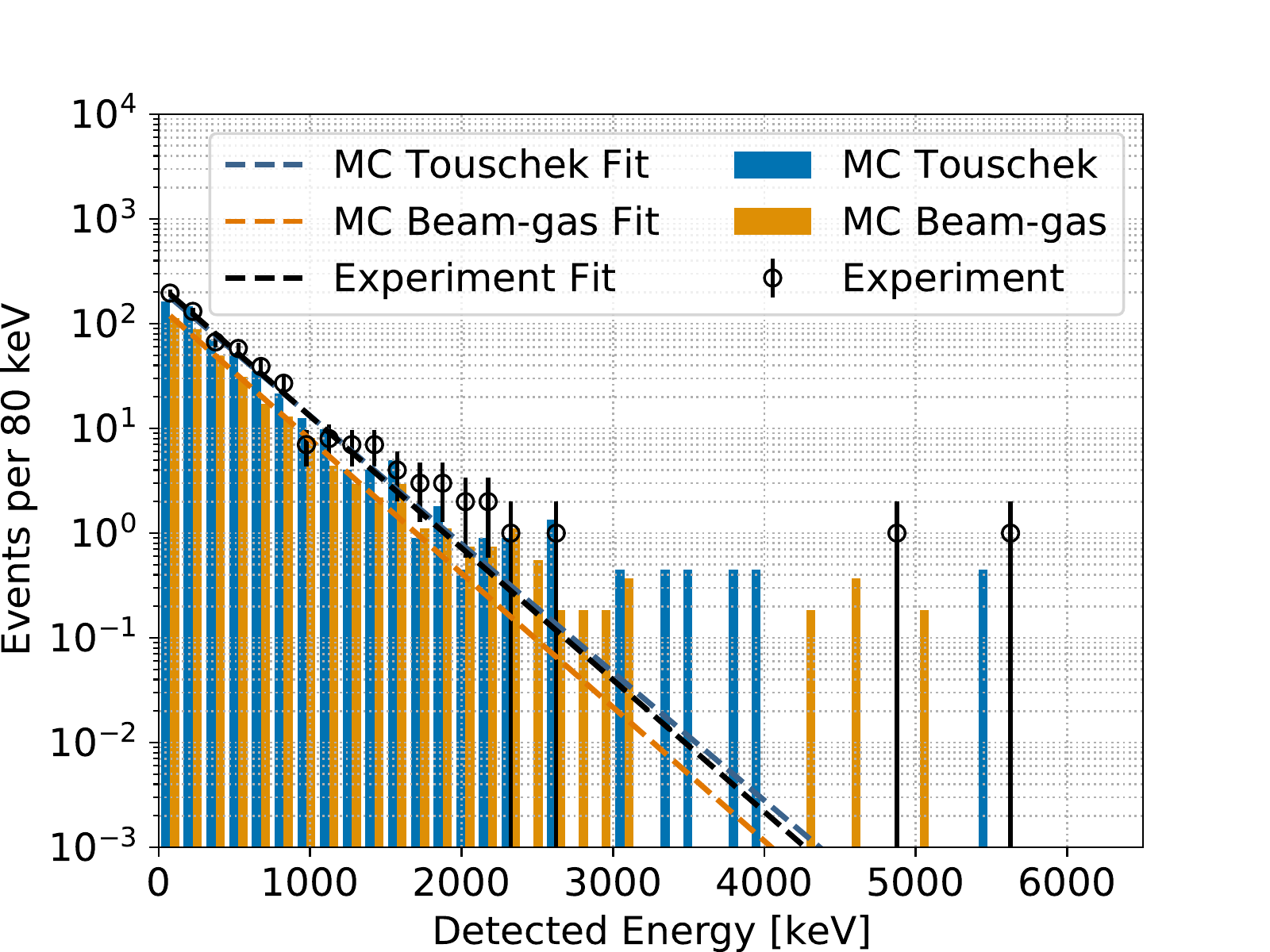}
\includegraphics[width=\columnwidth]{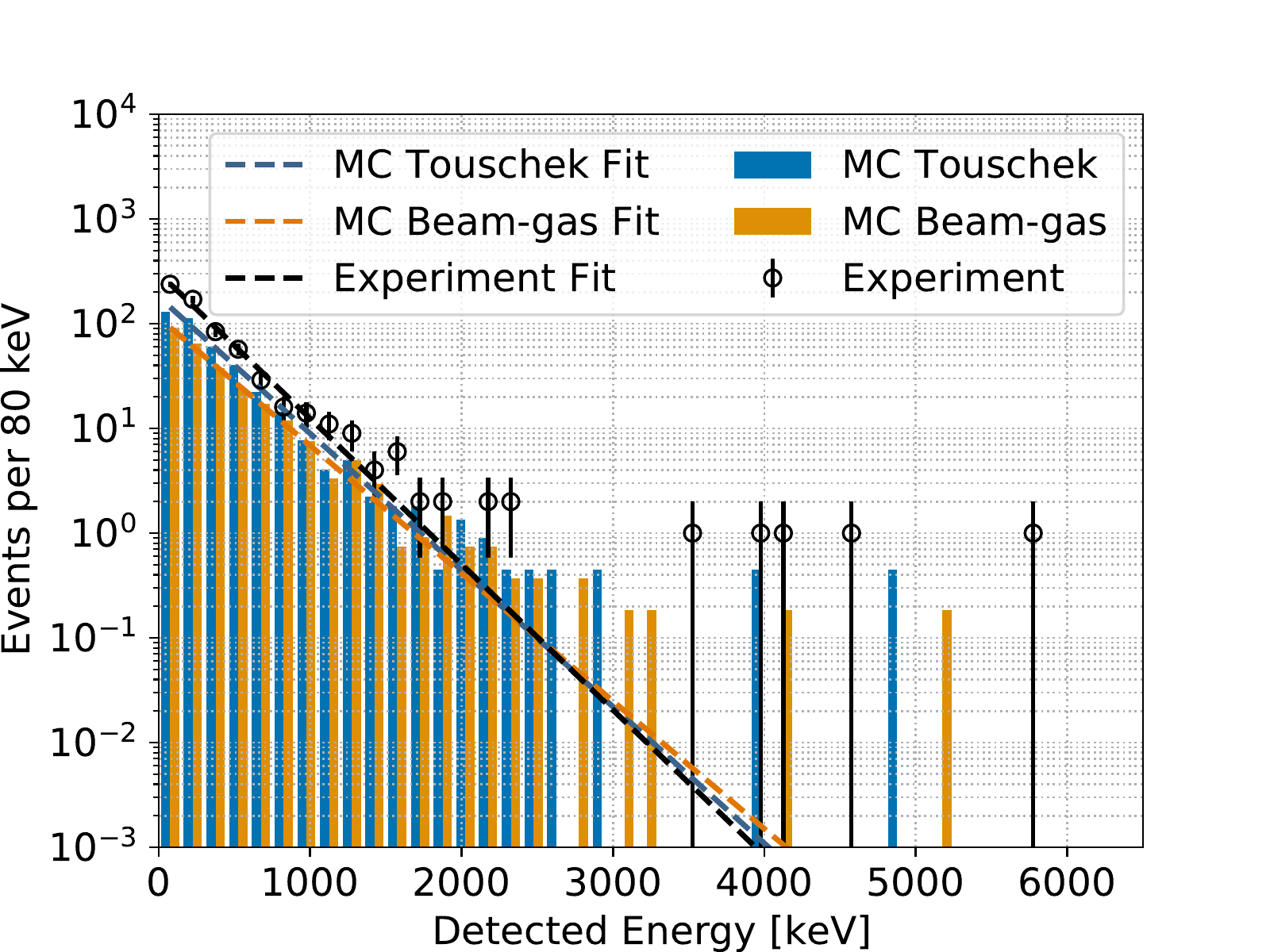}
\caption{Detected energy distribution for nuclear recoil candidates in TPC H (top) and TPC V (bottom) for the LER run.}
\label{fig:LER-E}
\end{figure}

Fig. \ref{fig:LER-E} shows the recoil energy distributions in experimental data and the reweighted Monte Carlo simulation for the LER run. The recoil energy distributions are fit with a decaying exponential of the form \(Ae^{-bE}\), where \(E\) is the recoil energy in keV$_{\rm ee}$. The fit results are shown in Table \ref{tbl:LER-E}. For parameter \(A\), we find that the simulation agrees with experimental data within errors for TPC V but overestimates the total measured background in TPC H. We also find that the spectral shapes of all distributions in each TPC---parameter \(b\) in the fit---are equivalent, suggesting that nuclear recoil production and detector materials are well modeled in the simulation even despite the noticeable disagreement in modeling carbon and oxygen recoils, as shown in Fig. \ref{fig:tpc-dEdx} and \ref{fig:tpc-EvsL}. However, we will require a different analysis to isolate beam-gas from Touschek backgrounds to further validate the simulation, which will be discussed in Sec. \ref{sec:org1a700f1}.

\begin{table}[htbp]
\caption{\label{tbl:LER-E}Results of fitting the recoil energy spectra for TPCs H and V for Monte Carlo and experimental data for the LER run.}
\centering
\begin{tabular}{rrrr}
TPC H & \(A\) & \(b\) & \(\chi\)\textsuperscript{2}/\(ndf\)\\
\hline
MC Beam-gas & 149 \textpm{} 12 & 0.0029 \textpm{} 0.0002 & 0.31\\
MC Touschek & 220 \textpm{} 14 & 0.0028 \textpm{} 0.0001 & 0.80\\
MC Total & 368 \textpm{} 18 & 0.0028 \textpm{} 0.0001 & 0.66\\
Exp. Data & 238 \textpm{} 16 & 0.0029 \textpm{} 0.0001 & 1.31\\
 &  &  & \\
TPC V &  &  & \\
\hline
MC Beam-gas & 114 \textpm{} 10 & 0.0028 \textpm{} 0.0002 & 0.25\\
MC Touschek & 180 \textpm{} 13 & 0.0030 \textpm{} 0.0002 & 0.58\\
MC Total & 293 \textpm{} 17 & 0.0029 \textpm{} 0.0001 & 0.44\\
Exp. Data & 306 \textpm{} 19 & 0.0032 \textpm{} 0.0002 & 1.21\\
\end{tabular}
\end{table}

\begin{figure}
\centering
\includegraphics[width=\columnwidth]{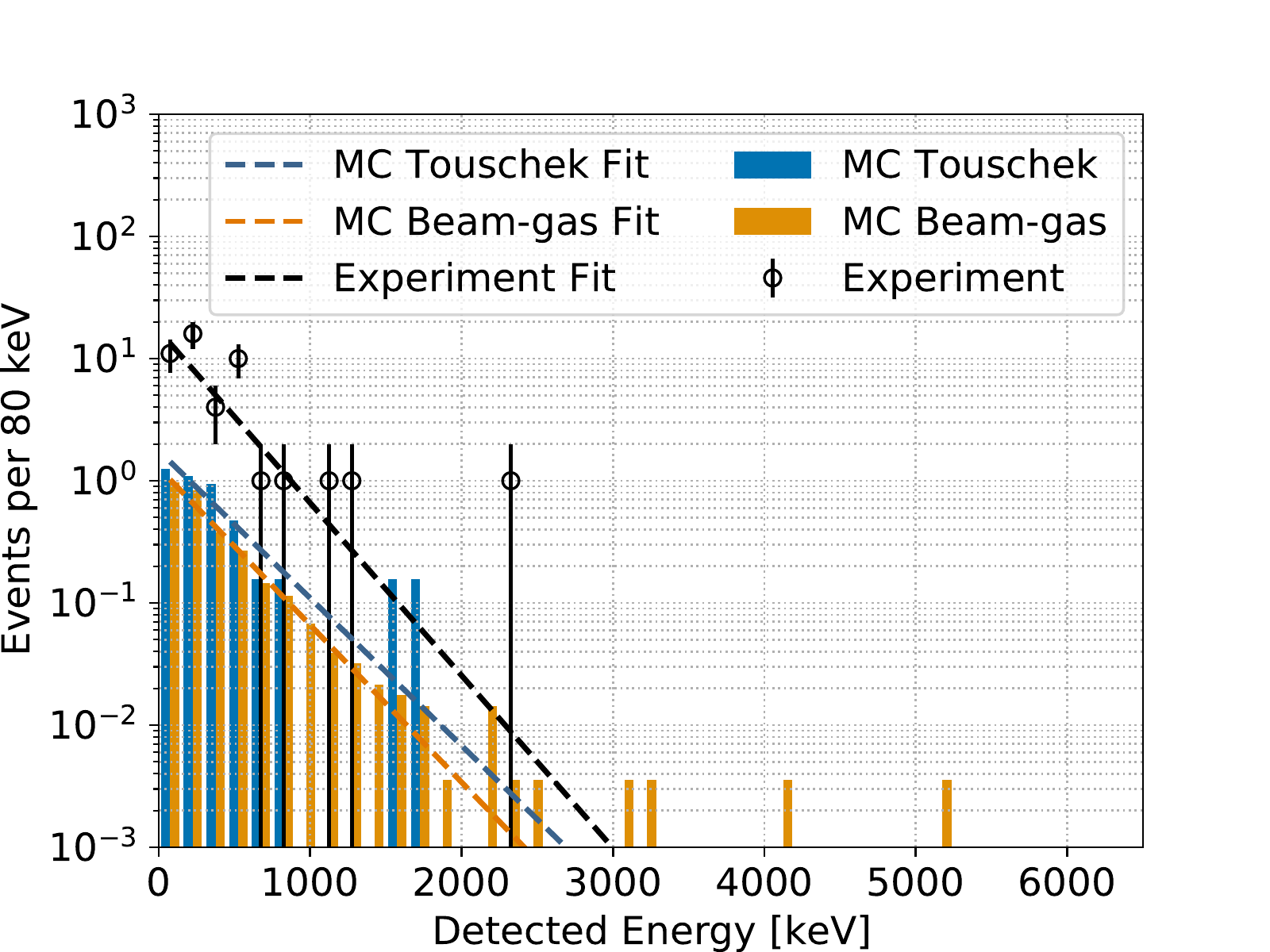}
\includegraphics[width=\columnwidth]{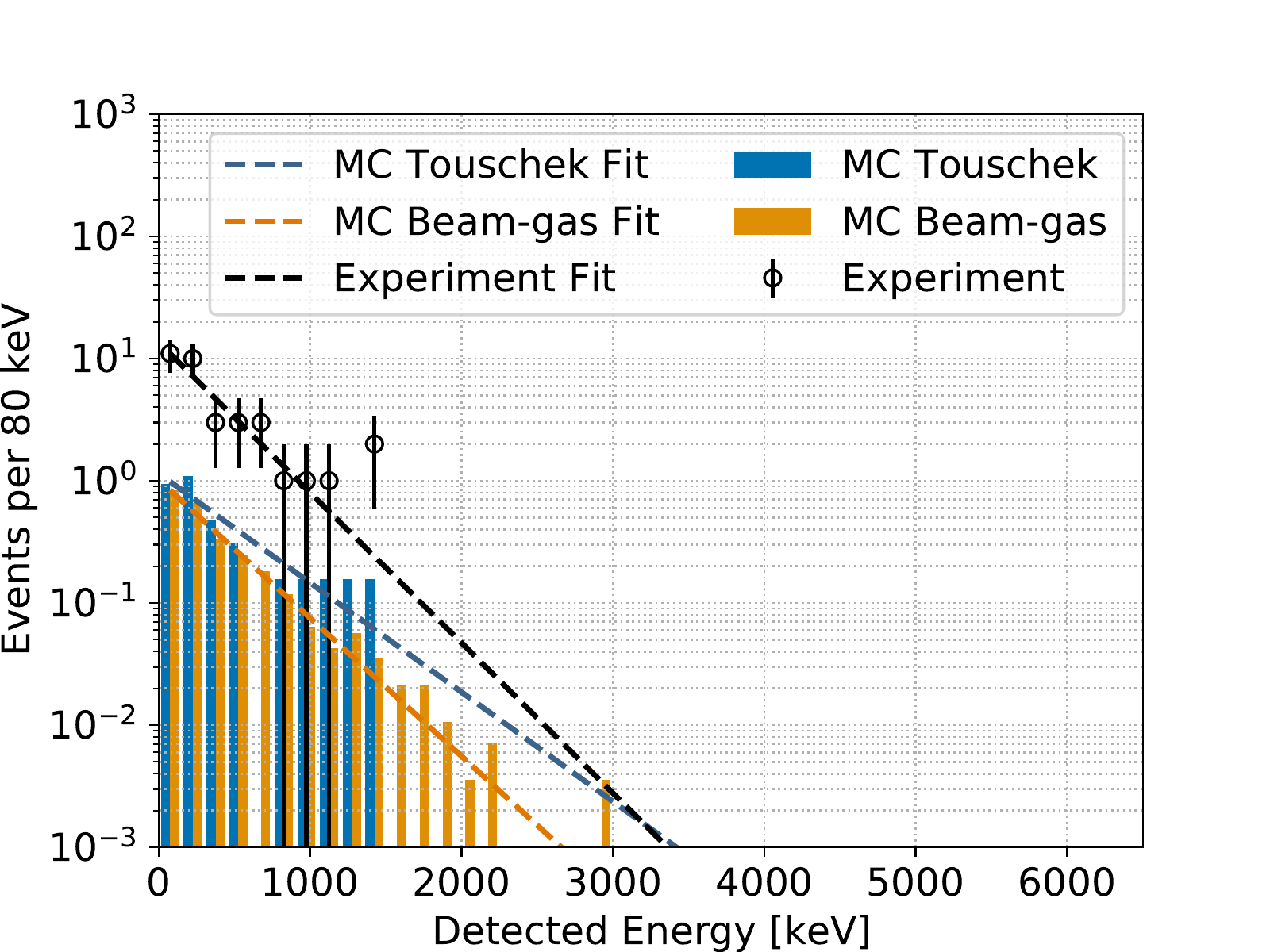}
\caption{Detected energy distribution for nuclear recoil candidates in TPC H (top) and TPC V (bottom) for the HER run.}
\label{fig:HER-E}
\end{figure}

We perform the same analysis for the HER run. We first calibrate the energy scale for the HER run period by following the procedure presented in Section \ref{sec:eCal}. The correction factors are shown in Table \ref{tbl:dQdx-calibrations-HER}. As can be seen, the gain is significantly lower in both TPCs during this run. This is due to the fact that the volumetric flow rate of the gas was set to approximately a factor of 5 lower than in the LER run. Finally, the energy recoil spectra for the HER run are shown in Fig. \ref{fig:HER-E}, and the fit results are shown in Table \ref{tbl:HER-E}. Here we find that the parameter \(b\) is again consistent within errors in both experimental and simulated data. Just as in the results from the LER run, we thus conclude that while validation of the simulation of the recoil energy spectra is an important result, we must utilize other methods to separate beam-gas from Touschek backgrounds.

\begin{table}[htbp]
\caption{\label{tbl:dQdx-calibrations-HER}Table of values of corrected $\mathrm{d}Q/\mathrm{d}x$ in TPC H, TPC V, and Monte Carlo simulation
and resulting conversion factors for the HER run.}
\centering
\begin{tabular}{rrr}
 & Average \(\mathrm{d}Q/\mathrm{d}x\) & Correction Factor\\
\hline
Simulation & 3227 & --\\
TPC H & 2172 & 1.49\\
TPC V & 1657 & 1.95\\
\end{tabular}
\end{table}

\begin{table}[htbp]
\caption{\label{tbl:HER-E}Results of fitting the recoil energy spectra for TPCs H and V for Monte Carlo and experimental
data for the HER run.}
\centering
\begin{tabular}{rrrr}
TPC H & \(A\) & \(b\) & \(\chi\)\textsuperscript{2}/\(ndf\)\\
\hline
MC Beam-gas & 1.3 \textpm{} 1.1 & 0.0029 \textpm{} 0.0019 & 0.01\\
MC Touschek & 1.8 \textpm{} 1.4 & 0.0028 \textpm{} 0.0020 & 0.09\\
MC Total & 3.04 \textpm{} 1.8 & 0.0028 \textpm{} 0.0016 & 0.03\\
Exp. Data & 17.0 \textpm{} 4.22 & 0.0032 \textpm{} 0.0006 & 1.8\\
 &  &  & \\
TPC V &  &  & \\
\hline
MC Beam-gas & 1.1 \textpm{} 1.0 & 0.0026 \textpm{} 0.0019 & 0.01\\
MC Touschek & 1.1 \textpm{} 1.1 & 0.0021 \textpm{} 0.0019 & 0.05\\
MC Total & 2.17 \textpm{} 1.48 & 0.0023 \textpm{} 0.0014 & 0.02\\
Exp. Data & 13.6 \textpm{} 4.0 & 0.0028 \textpm{} 0.0007 & 0.59\\
\end{tabular}
\end{table}

\subsection{Nuclear recoil event rate versus beam size}
\label{sec:org1a700f1}
Detected event rates from beam-gas and Touschek are expected to change with beam parameters. As such, we can measure the nuclear recoil event rate while varying the appropriate accelerator conditions and compare the measurement to the rate predicted from simulated data. Ref.~\cite{lewis19_first_measur_beam_backg_at_super} shows that the rate due to beam-gas scattering should linearly increase with \(IPZ_e^2\), where \(I, P,\) and \(Z_e\) are the beam current, pressure, and effective \(Z\) of the residual gas in the beam;  and Touschek backgrounds should linearly increase with \(I^2/\sigma_y\), where \(\sigma_y\) is the beam size in the \(y\) direction as measured by the SuperKEKB X-ray monitors \cite{Flanagan:2005}. Given the hours-long duration of the TPC runs, we choose \(I, P, Z_e\), and \(\sigma_y\) such that they correspond to the average values measured along the entirety of the LER during the duration of individual measurements, and we denote these values as \(\overline{I}\), \(\overline{P}\), \(\overline{Z_e}\), and \(\overline{\sigma_y}\). We then describe the \emph{sensitivity} of the TPCs to beam-gas, \(S_{bg}\), and Touschek, \(S_T\), backgrounds by:

\begin{equation}
\label{eq:heuristic}
\frac{R}{\bar{I}\bar{P}\overline{Z_e}^2} = S_{bg} + S_T\frac{\bar{I}}{\bar{P}\overline{\sigma_y}\overline{Z_e}^2}
\end{equation}
This provides a description of the expected rate of detected nuclear recoil events, \(R\), versus the inverse of the beam size,
\(1/\overline{\sigma{}_y}\) while controlling for other changing beam conditions like pressure, current, and gas composition. Furthermore, it allows for a simple
linear fit to obtain $R$, the observed rate of nuclear recoils in the TPCs, versus \(1/\overline{\sigma{}_y}\) to separate beam-gas and Touschek
backgrounds.

\begin{figure}
\centering
\includegraphics[width=\columnwidth]{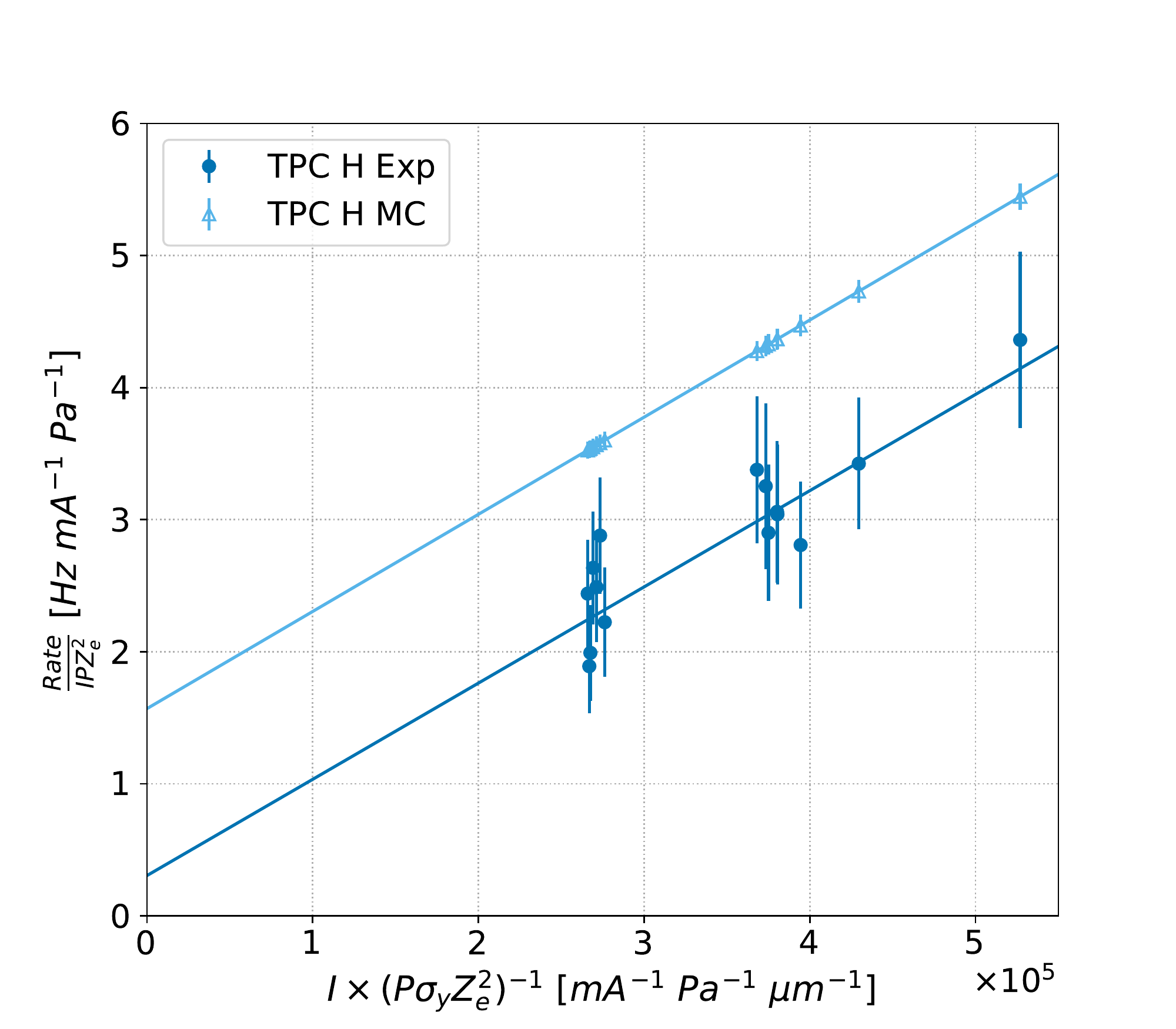}
\includegraphics[width=\columnwidth]{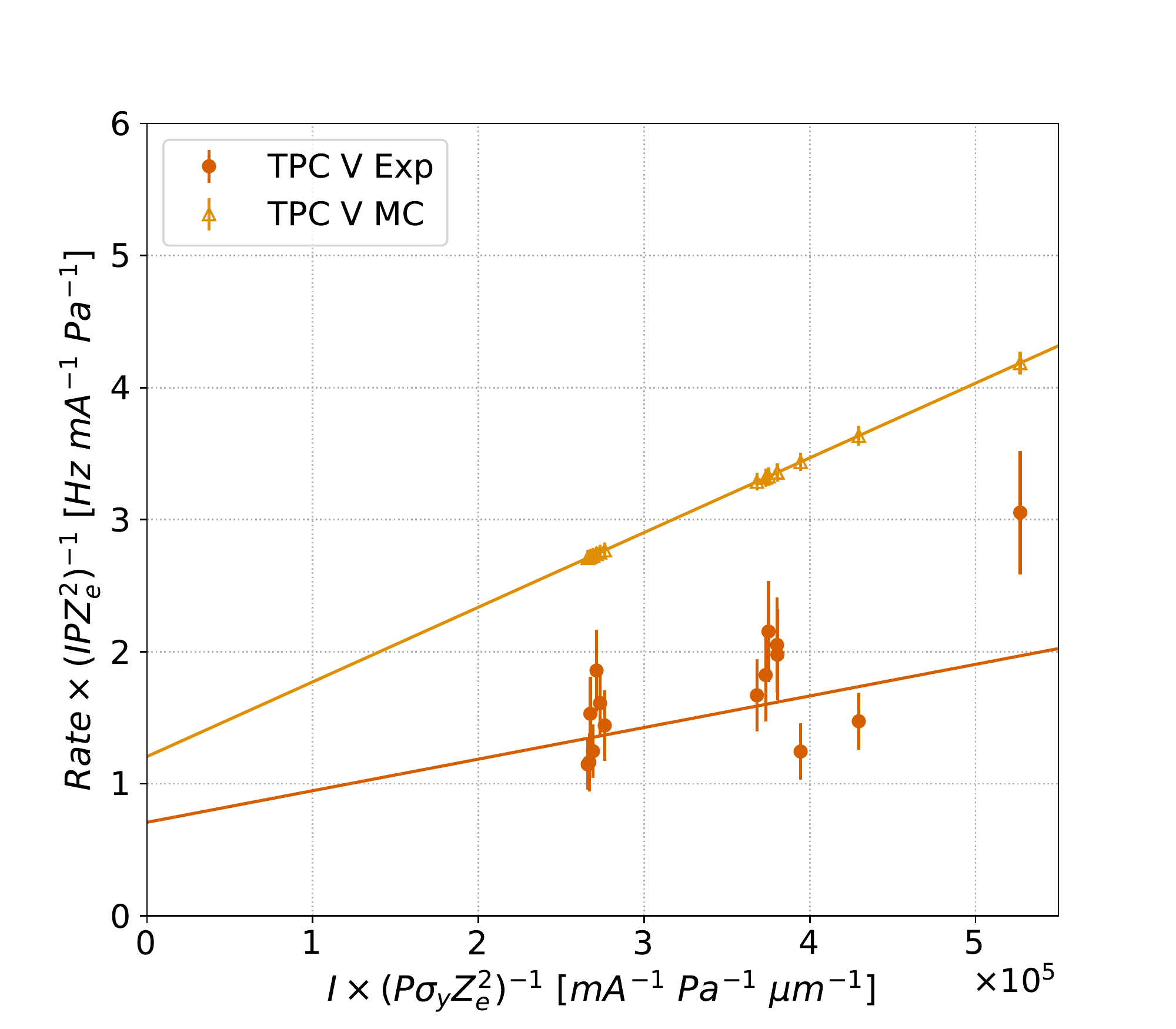}
\caption{Plot of the fast neutron rates in TPC H (top) and TPC V (bottom) during the LER run.}
\label{fig:tpc-sensitivities}
\end{figure}

The measured nuclear recoil rates in the TPCs versus LER beam size are shown in Fig. \ref{fig:tpc-sensitivities}. The obtained sensitivities can be integrated to directly obtain the measured and predicted rates to give a yield, denoted as \(N_T\) for the yield of Touschek events and \(N_{bg}\) for beam-gas events. The observed yields are shown in Table \ref{tbl:heuristic}. We find agreement in \(N_T\) for TPC H, but substantial disagreement for \(N_T\) in TPC V. Furthermore, the simulation overestimates \(N_{bg}\) by nearly a factor of two in TPC V and a factor of three in TPC H.

\begin{table}[htbp]
\caption{\label{tbl:heuristic}Calculated yield from the measured rates of nuclear recoils from beam-gas and Touschek backgrounds
shown in Fig. \ref{fig:tpc-sensitivities} for both experimental data and Monte Carlo in each TPC.}
\centering
\begin{tabular}{rrr}
TPC H & \(N_{bg}\) & \(N_T\)\\
\hline
MC & 339 \textpm{} 19 & 580 \textpm{} 21\\
Exp. & 66 \textpm{} 22 & 574 \textpm{} 25\\
 &  & \\
TPC V &  & \\
\hline
MC & 261 \textpm{} 17 & 445 \textpm{} 19\\
Exp. & 153 \textpm{} 12 & 188 \textpm{} 13\\
\end{tabular}
\end{table}

\subsection{Directional analysis of nuclear recoils}
\label{sec:org48444a6}
Next, we seek to provide directional measurements of detected nuclear recoils. We do so by first discriminating helium recoil events that originate from the direction of the beam-line from neutron events originating elsewhere. Secondly, we will attempt to fit for the fractional contribution of Touschek and beam-gas events within the angular distribution of events in experimental data.

In each TPC, we select events with an axial track fit along an axis between the TPC and the beam-pipe. Utilizing the head-charge fraction (HCF) variable, we fit for the number of events with positive radial velocity---specifically events with $\mathrm{d}r/\mathrm{d}t > 0$ in the Belle II coordinate system, with $r = 0$ corresponding to the beam-line---and the number of events with radial velocity in the opposite direction. The templates for these events are built from histograms of the truth HCF distributions of simulated helium recoil events. Using these templates, we then fit for the yield of each template to the histogram of experimentally measured HCF distributions.

To do so, we impose a \emph{positive hypothesis} on all events, wherein we assign the values of \(\theta~\rm and~\phi\) to all events such that they have a positive radial velocity with respect to the Belle II coordinate system. In TPC H, this corresponds to a \(\phi\) for all detected events in the range of \(-90^{\circ} < \phi < 90^{\circ}\). The axis connecting the beam-line and TPC H, or \emph{Line-of-Sight} (\emph{LoS}), falls along the \(-x\mathrm{-axis}\), corresponding to an angle of \(\phi{} = 180^{\circ}\) and \(\theta = 90^{\circ}\) in Belle II coordinates. In TPC V, this corresponds to a \(\phi\) for all detected events such that \(-180^{\circ} < \phi < 0^{\circ}\) with the \emph{LoS} of TPC V falling along the \(-y\mathrm{-axis}\). We define an event acceptance of \(\pm20^{\circ}\) in \(\phi\) from the \emph{LoS}. This corresponds of \(160^{\circ} < |\phi| < 180^{\circ}\) in TPC H and -\(70^{\circ} < \phi < -110^{\circ}\) in TPC V. These selections and the resulting cumulative efficiencies are shown in Table \ref{tbl:all-sel-effs} for both TPCs.

\begin{figure}
\centering
\includegraphics[width=\columnwidth]{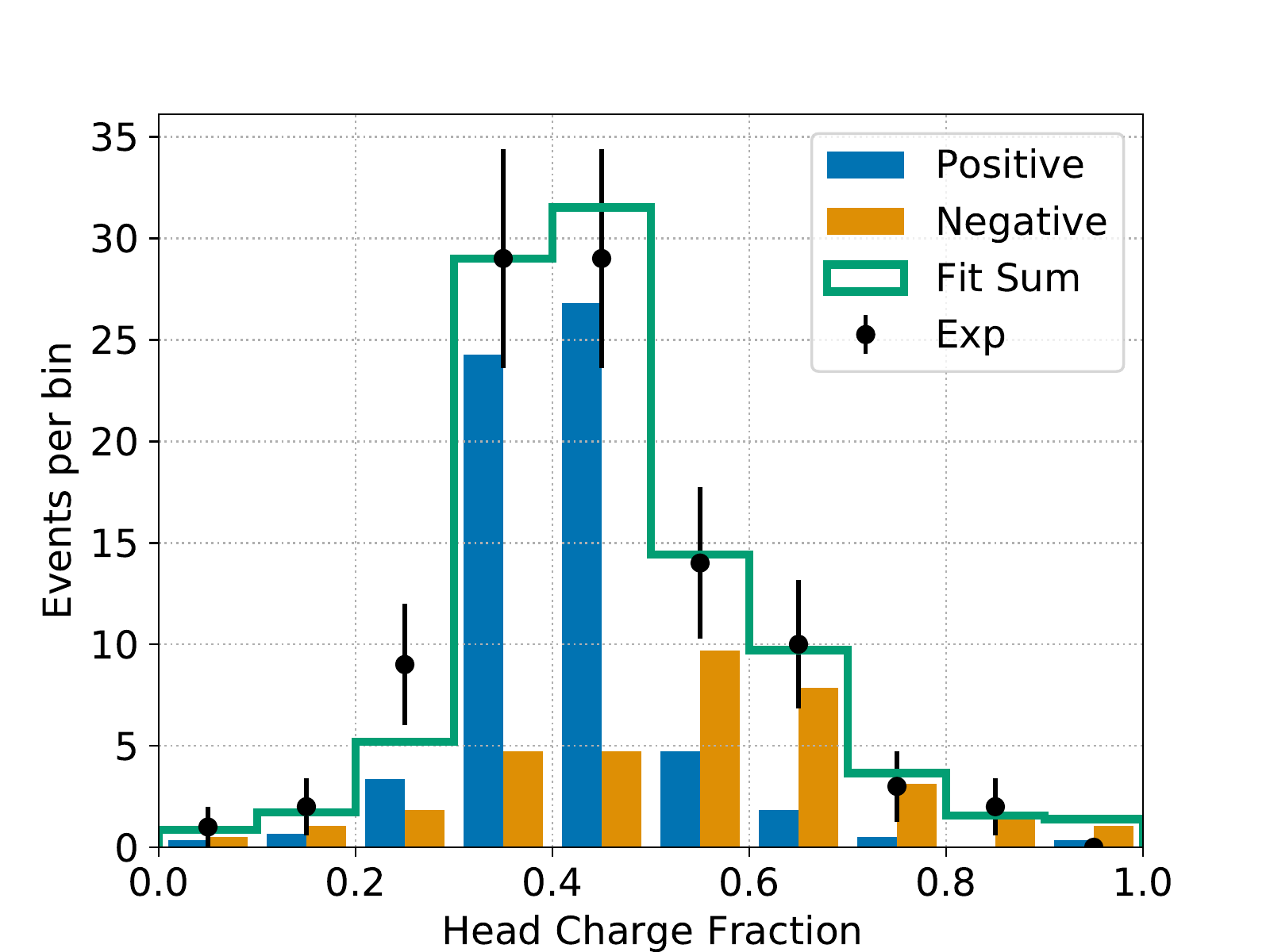}
\caption{Fitted distribution of fractional charge for selected nuclear recoils in TPC H in experimental data using templates derived from Monte Carlo for events with positive or negative radial velocity.}
\label{fig:headtail-tpcH}
\end{figure}

\begin{figure}
\centering
\includegraphics[width=\columnwidth]{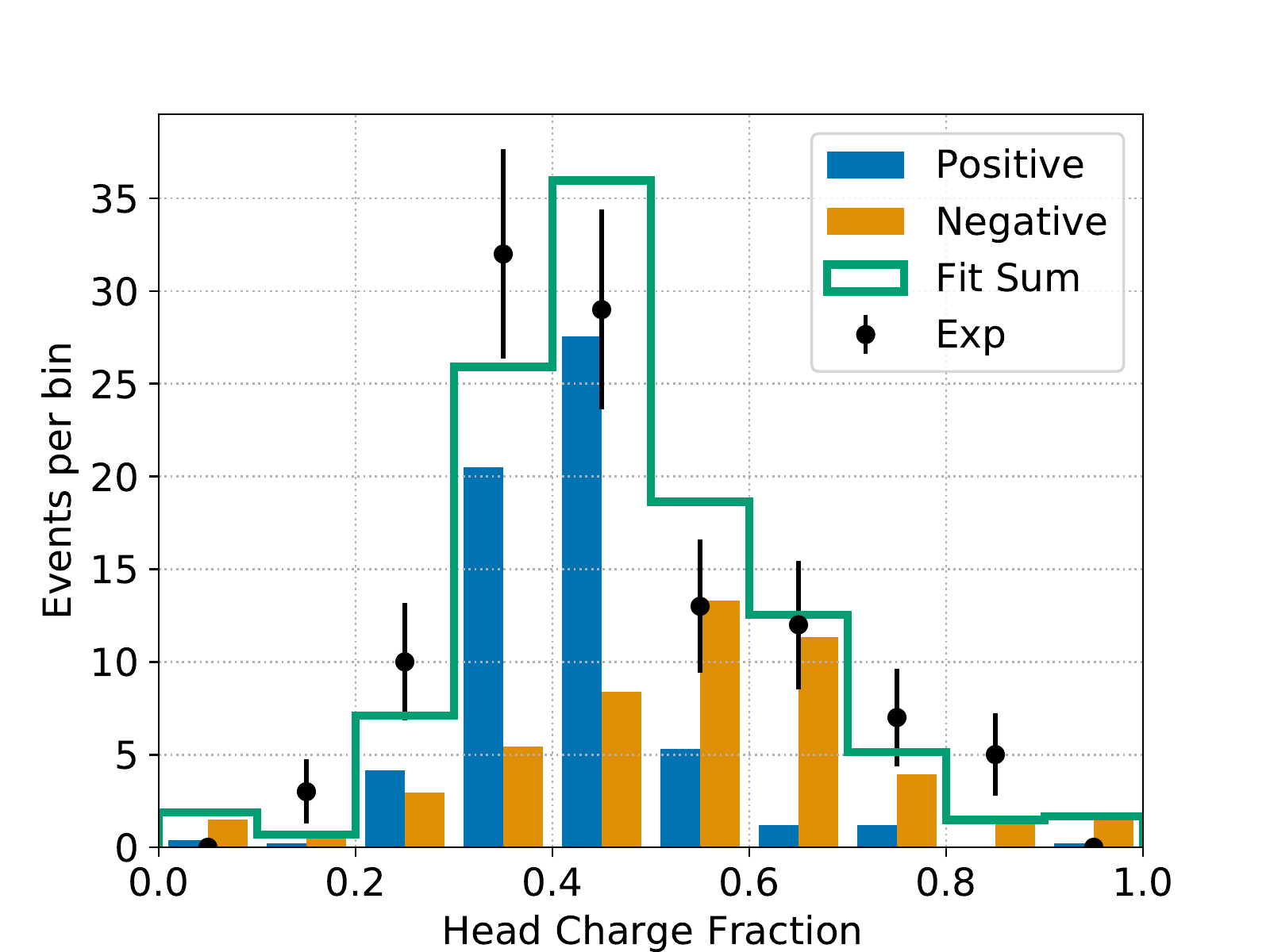}
\caption{Fitted distribution of fractional charge for selected nuclear recoils in TPC V in experimental data using templates derived from Monte Carlo for events with positive or negative radial velocity.}
\label{fig:headtail-tpcV}
\end{figure}

The results of the analysis are shown in Figures \ref{fig:headtail-tpcH} and \ref{fig:headtail-tpcV} for TPCs H and V, respectively. The template histograms are scaled to the fitted fractional composition of each event type as given by the fit. The green line shows the sum of the scaled templates in each bin. The fitted fractional yields are given in Table \ref{tbl:head-tail-fit}. In TPC H, we find the fitted fractional yields are equivalent to the prediction from Monte Carlo, within errors, at a composition of 75\% positive to 25\% negative travelling recoils. For TPC V, the Monte Carlo also predicts 75\% positive to 25\% negative events in TPC V, but the fits to experimental data show 50\% of each component at \(\sim 1.9\sigma\). One possible explanation could be that the component of events with negative radial velocity is larger in experimental data, but only seen in the vertical plane of the SuperKEKB beam-line. A future study with higher statistics samples of experimental and simulated neutron signal events and electron background simulations could provide further insights into this effect.

\begin{table}[htbp]
\caption{\label{tbl:head-tail-fit}Fraction of events with positive and negative radial velocity predicted in simulation and from fitting yields to
simulated and experimental data in TPCs H and V.  \(N_{+}\) corresponds to the fraction of events with positive radial velocity and \(N_{-}\)
corresponds to the fraction of events with negative radial velocity. Errors on the truth values correspond to the square-root of the number of generated events, whereas remaining errors are the errors obtained from the log-likelihood fit.}
\centering
\begin{tabular}{rrr}
TPC H & \(N_{+}\) & \(N_{-}\)\\
\hline
Truth & 0.73 \textpm{} 0.04 & 0.27 \textpm{} 0.02\\
MC Fit & 0.73 \textpm{} 0.05 & 0.27 \textpm{} 0.04\\
Exp Fit & 0.64 \textpm{} 0.10 & 0.36 \textpm{} 0.09\\
 &  & \\
TPC V &  & \\
\hline
Truth & 0.75 \textpm{} 0.06 & 0.25 \textpm{} 0.02\\
MC Fit & 0.75 \textpm{} 0.05 & 0.25 \textpm{} 0.04\\
Exp Fit & 0.52 \textpm{} 0.09 & 0.47 \textpm{} 0.09\\
\end{tabular}
\end{table}

Next, we attempt to fit distributions of \(\cos\theta\) in experimental data to templates obtained from simulated recoils from beam-gas and Touschek backgrounds. The angle \(\theta\) for both TPCs corresponds to a location along the SuperKEKB beam-line. For this analysis, we select helium recoils with positive radial velocity in both TPCs by selecting events with HCF < 0.5, meaning that we use all selections listed in Table \ref{tbl:all-sel-effs}. As can be seen, the remaining number of simulated events after applying all of these selections is small, thereby introducing uncertainty from Poisson statistics in the individual bins of the template histograms. To ensure we can proceed, we must first determine whether the distributions of \(\cos\theta\) in the simulated beam-gas and Touschek distributions can be separated. We do so by performing a Kolmogorov-Smirnov test on the two components to calculate the probability that the distributions are consistent with each other. We find that \(p = 0.620 \) in TPC H and \(p = 0.028\) in TPC V. The low probability in TPC V suggests, but does not prove, that Touschek and beam-gas events can be disentangled by the shape of the \(\cos\theta\) distributions alone. Given the low statistics, we use the TFractionFitter class to fit the experimental data with the Monte Carlo templates, as we cannot ignore the statistical uncertainties in the simulated data.

\begin{figure}[h]
\includegraphics[width=\columnwidth]{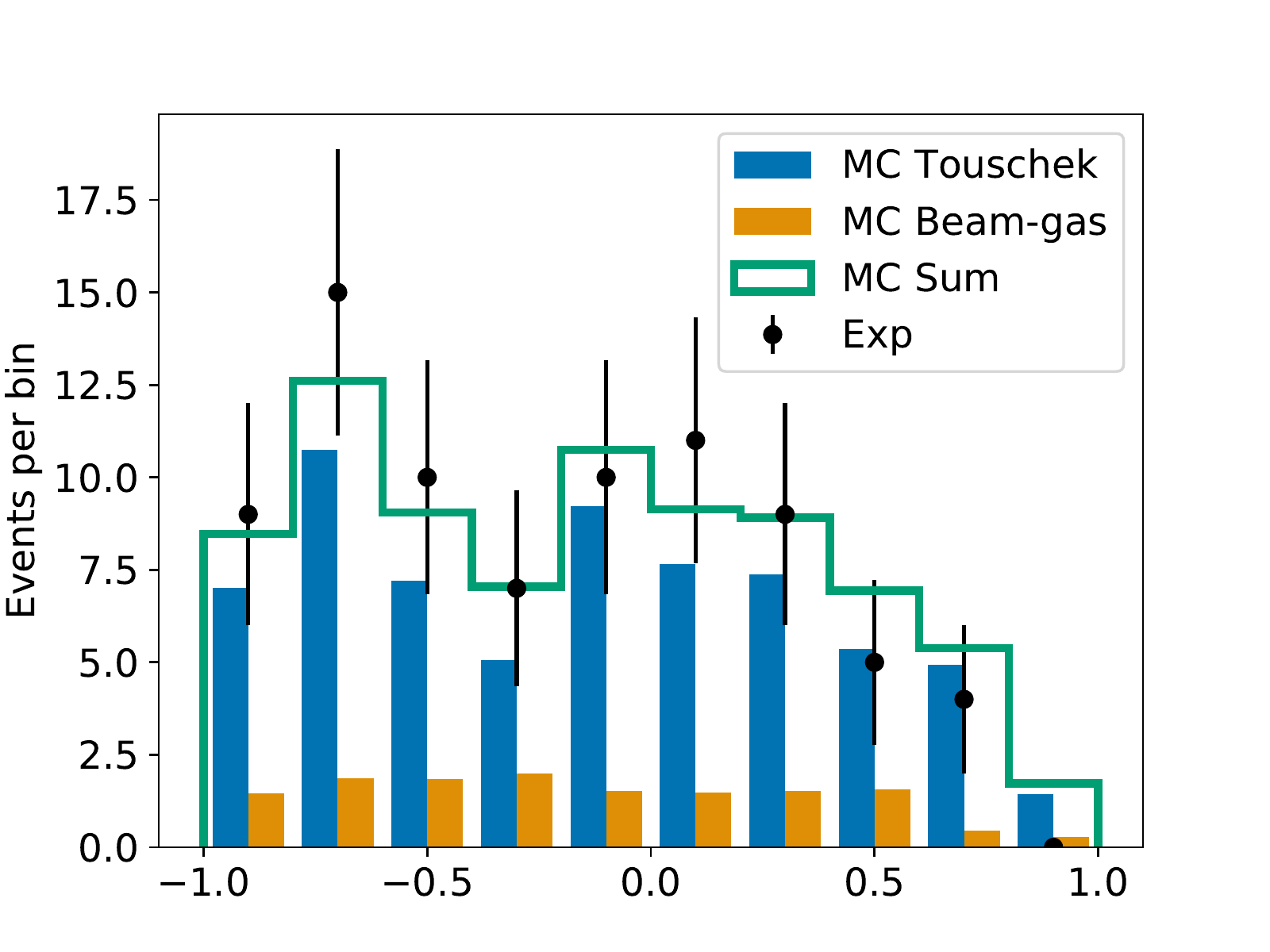}
\includegraphics[width=\columnwidth]{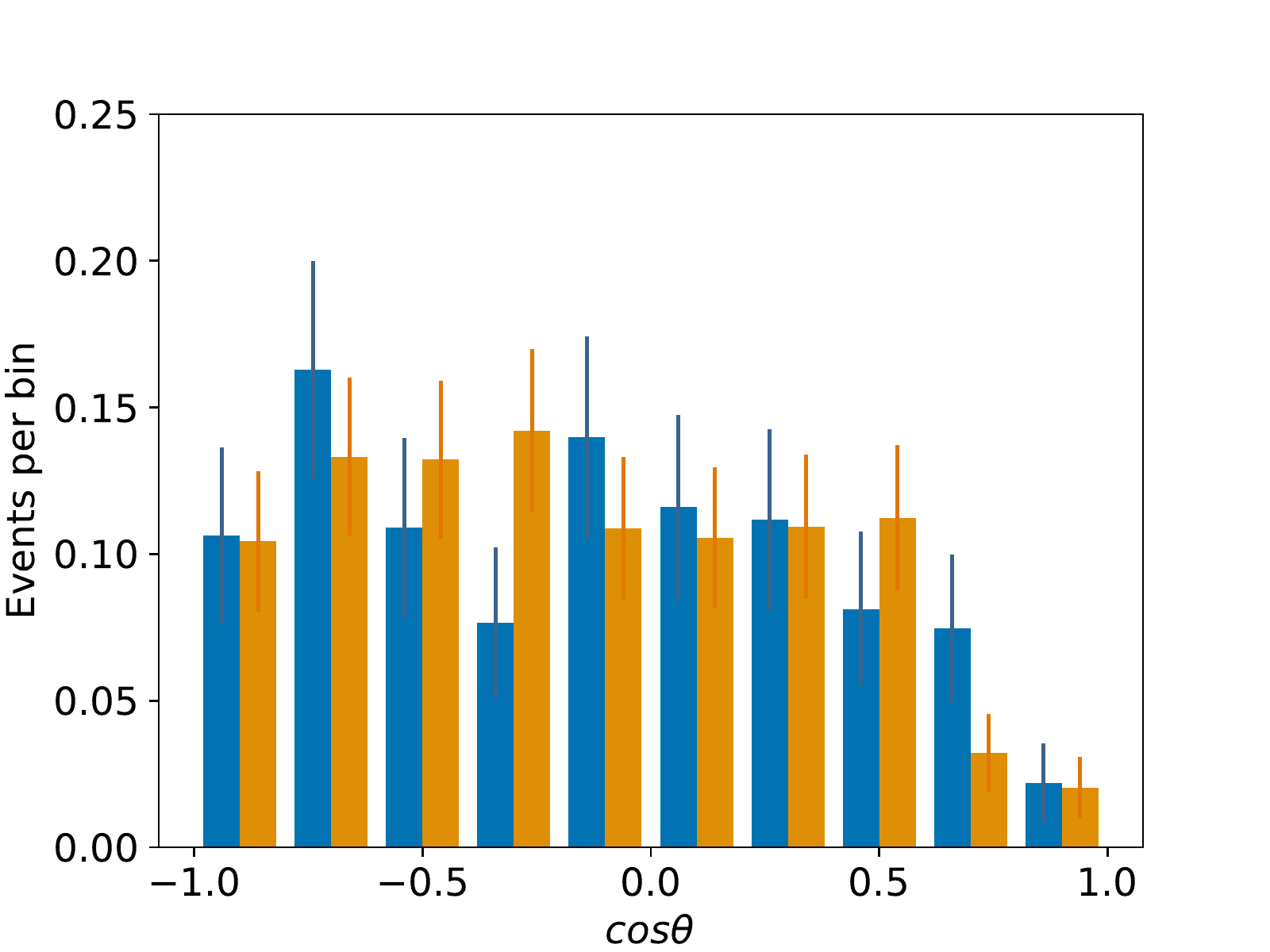}
\caption{(Top) Fitted distribution of $\cos\theta$ in experimental data in TPC H (black points) with fractional yields of Touschek (blue)
and beam-gas (orange) events in simulated data.  The green line corresponds to the sum of the templates.
(Bottom) Normalized templates used for fitting to the black points including Poisson uncertainties in each bin.}
\label{fig:cosTheta-tpcH}
\end{figure}

\begin{figure}[h]
\includegraphics[width=\columnwidth]{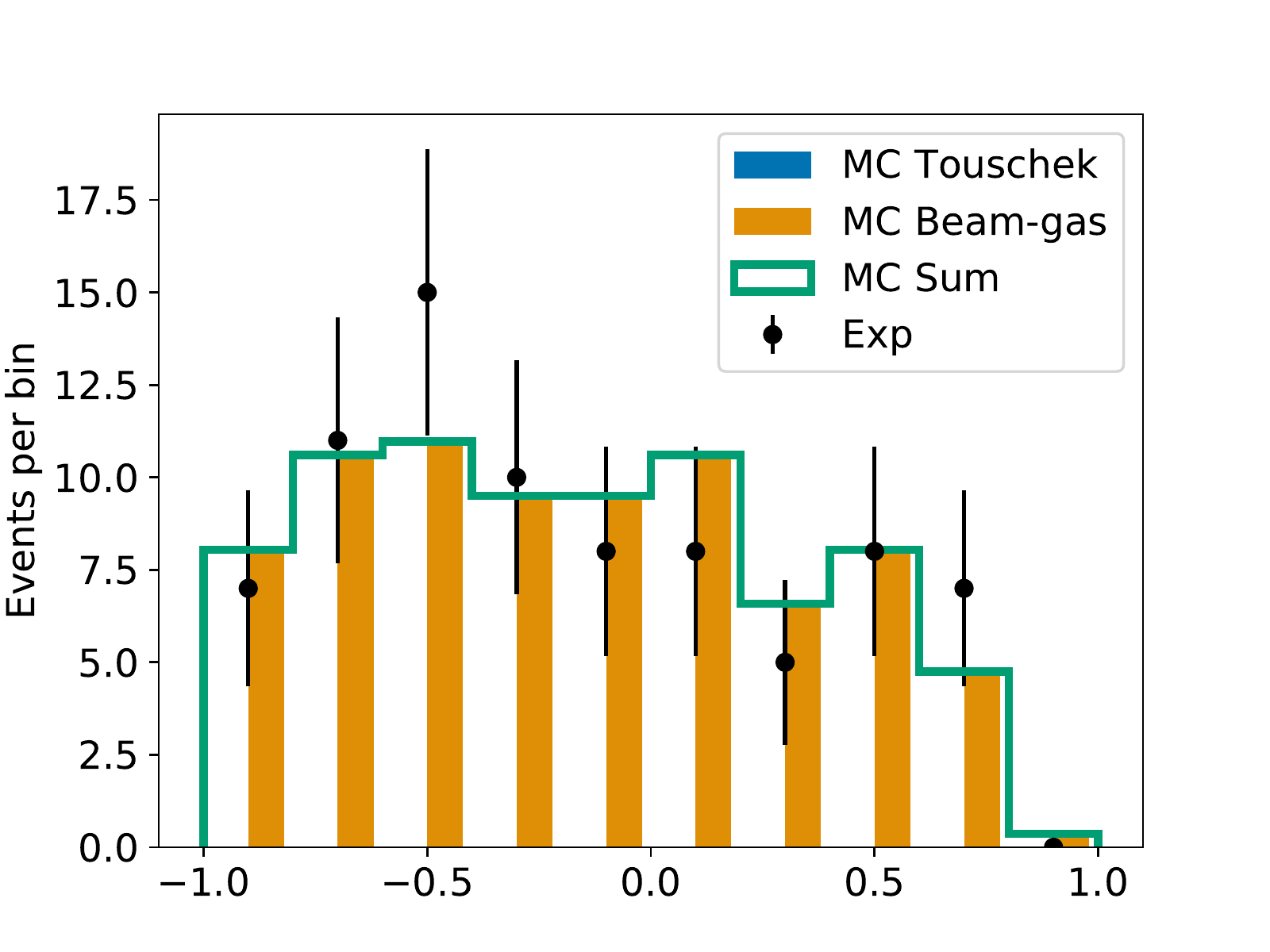}
\includegraphics[width=\columnwidth]{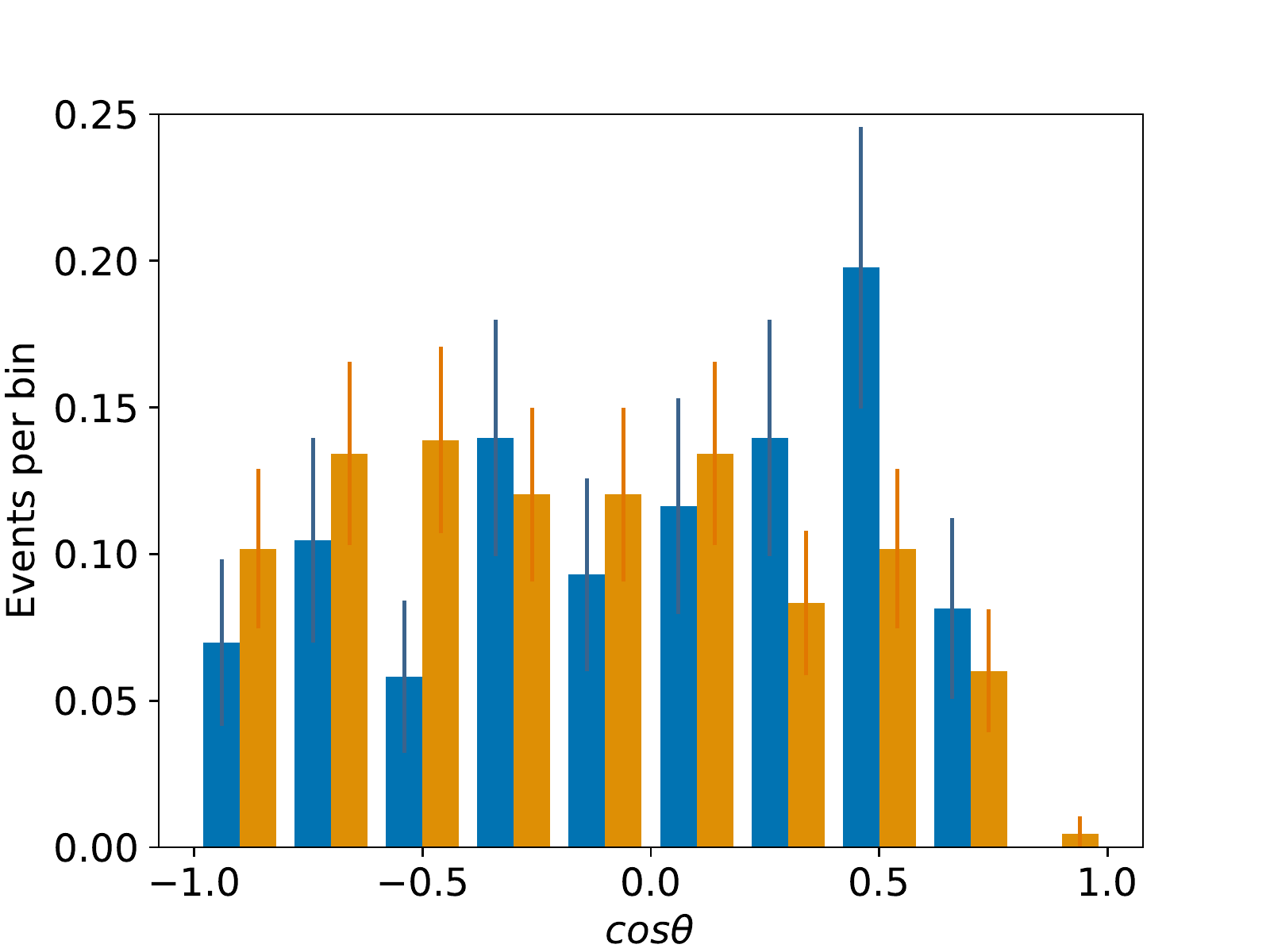}
\caption{(Top) Fitted distribution of $\cos\theta$ in experimental data in TPC V (black points) with fractional yields of Touschek (blue)
and beam-gas (orange) events in simulated data.  The green line corresponds to the sum of the templates. (Bottom) Normalized
templates used for fitting to the black points including Poisson uncertainties in each bin.}
\label{fig:cosTheta-tpcV}
\end{figure}

The templates and result of the fits are shown in Fig. \ref{fig:cosTheta-tpcH} and \ref{fig:cosTheta-tpcV} for TPCs H and V, respectively. The fitted fractional compositions are shown in Table \ref{tbl:cosTheta-fits} and are compared to the results of the beam emittance scan analysis from Section \ref{sec:org1a700f1}. As can be seen, the results of both methods are consistent. Furthermore, both TPCs measure a higher rate in \(\cos\theta{}\) < 0. This region is in the downstream direction of the LER beam. Considering that these data samples are taken from only the LER beam, this indicates that we are likely seeing forward showers from the LER interacting with the beam-pipe downstream of the TPCs with respect to the LER beam-direction.

While these results are clearly statistics limited, this method demonstrates a possible separation of Touschek backgrounds from beam-gas backgrounds without the need for time-consuming, dedicated experimental runs where accelerator parameters are systematically varied. In principle, should this analysis method be verified, this could allow for analysis of fast neutron backgrounds symbiotic with normal Belle II operation, thereby eliminating the need for dedicated long-duration neutron runs.

\begin{table}[htbp]
\caption{\label{tbl:cosTheta-fits}Calculated fractions of Touschek and beam-gas events from fits of \(\cos\theta{}\) in both TPCs, as shown
in Fig. \ref{fig:cosTheta-tpcH} and \ref{fig:cosTheta-tpcV}, compared to calculated yields of Touschek and beam-gas events from the beam emittance scan analysis.}
\centering
\begin{tabular}{rrr}
TPC H & \(N_{bg}\) & \(N_T\)\\
\hline
\(\cos\theta{}\) & \(0.18 \pm ^{0.53}_{0.18}\) & \(0.83 \pm ^{0.16}_{0.83}\)\\
Heuristic & 0.10 \textpm{} 0.03 & 0.90 \textpm{} 0.04\\
 &  & \\
TPC V &  & \\
\hline
\(\cos\theta{}\) & \(1.0 \pm ^{0.0}_{0.78}\) & \(0.00 \pm ^{0.66}_{0.00}\)\\
Heuristic & 0.45 \textpm{} 0.04 & 0.55 \textpm{} 0.04\\
\end{tabular}
\end{table}

\section{Summary and discussion}
The results presented herein have important implications for two distinct topics. Firstly, the TPC system development, deployment, and simulation campaign have resulted in detailed rate, energy, and directional measurements of beam-induced fast neutron recoils, providing a first validation of the neutron background simulation at SuperKEKB. Secondly, the performance of these TPCs merits discussion of their potential more broadly in low background experiments.

\subsection{Analysis of fast neutron backgrounds at SuperKEKB}
The simulation and experimental campaign to measure fast neutron backgrounds at SuperKEKB has been a remarkable success. Using the performance of the TPC system, we have provided measurements of neutron-induced nuclear recoil rates, energy spectra, and directional composition as well isolated beam-gas from Touschek backgrounds. Comparison of these measurements to the SuperKEKB beam-loss simulation campaign has resulted in a promising outcome. The disagreement between experimental and simulated data for all of these measurements is simultaneously small enough to suggest that beam-losses at SuperKEKB are well modeled, yet large enough to merit targeted studies of these effects in the future through larger campaigns with more experimental and simulated data.

We note the following results from our fast neutron analyses. The LER simulations slightly overestimate the measured total of nuclear recoil events in TPC V and significantly overestimate the measured total nuclear recoils in TPC H.  The recoil energy spectral shapes agree remarkably well in simulated and experimental data. The coefficient of the recoil energy spectra---parameter \(A\) in the fit--- of simulated and experimental agree in TPC V but disagree in TPC H by a factor of 1.5, thereby warranting further study with more data. The Monte Carlo prediction and the experimental measurement agree that the fractional amount of events in TPC H with negative radial velocity is 25\% of the total yield. However, in TPC V, we measure the yield of these events to be 50\% at a significance of $\sigma = 1.9$, warranting future study with more data. The distributions of observed events versus $\cos\theta$ are consistent with simulation in both TPCs, indicating that predicted neutron production points along the beam-line near the TPCs for the LER match the observed rates. Lastly, we have presented a new analysis method for discriminating Touschek backgrounds from beam-gas backgrounds using vector directional measurements of nuclear recoils. The results of this method, using $\cos\theta$, are consistent with the results of the beam emittance scan shown in Ref.~\cite{lewis19_first_measur_beam_backg_at_super}, albeit with large uncertainties. Validation of this analysis technique with larger statistics could provide a method of analyzing fast neutron backgrounds at Belle II without the need for time-consuming machine studies requiring systematic variation of beam parameters, thereby eliminating the need to interrupt other operations and allowing symbiotic operation alongside Belle II data-taking. We also note that a more sophisticated analysis combining the beam size dependence, relative rates between TPCs, and the angular information of each TPC could have the best sensitivity for analysis beam background induced fast neutrons at SuperKEKB.

\subsection{TPC system performance comparisons}
Finally, we note that the high definition nuclear recoil imaging demonstrated here is also of interest for low background experiments. In directional dark matter searches, for example, the major challenges are to obtain good recoil angle resolution, good head-tail efficiency, and high electron rejection at the keV-scale recoil energies expected from dark matter recoils. Reviews of such efforts and ideas on possible future directions for the field can be found in Refs.~\cite{Vahsen:2020pzb, Mayet:2016zxu, Battat:2016pap, Vahsen:2021gnb, ahlen}.

To date, full 3D vector tracking of nuclear recoils has only been demonstrated by NEWAGE \cite{Yakabe:2020rua} using low pressure CF$_4$ gas. Others, such as MIMAC \cite{Santos:2011kf} and DMTPC \cite{Leyton:2016nit} have demonstrated head/tail discrimination in 2D using low pressure flourine targets. This makes the results presented here the first 3D vector tracking of helium recoils at atmospheric pressure. Although the TPCs used in the present work were optimized for directional detection of higher-energy fast neutron recoils, surprisingly, the performance presented here appears to exceed that of these existing directional dark matter detectors. For example, the angular resolution at fixed recoil energy reported here is lower, and our head/tail efficiency is higher, despite operating at atmospheric pressure, where the ratio of recoil length to diffusion is reduced, so that angular resolution degrades, compared to lower pressures.

These improvements result from a combination of five factors. First, our detectors are small, minimizing drift length and hence diffusion. Second, we utilize a helium target nucleus, which will travel further and with less scattering than a fluorine nucleus at a given recoil energy. Third, our gas is helium rich and thus has lower density at a given pressure, further extending recoil lengths. Fourth, the pixel ASIC charge readout has an extremely low noise level compared to the gain in the double GEM amplification stage, resulting in very high signal-to-noise, as seen in Fig. \ref{fig:evt-display}. Finally, the high readout segmentation helps with accurate recoil reconstruction and background rejection. The drawback of pixel ASIC readout, however, is relatively high cost and potentially high intrinsic radioactivity. A first cost-benefit analysis suggests that for larger dark matter detectors, 2D strip readout may be a better trade-off~\cite{Vahsen:2020pzb}. However, for precision studies of nuclear recoils and experiments at the lowest energies, pixel readout may remain the best choice. We plan to study these two development paths next. The excellent electron rejection demonstrated here is also of interest for low background experiments. We are currently investigating how this can be improved further with more sophisticated algorithms \cite{Ghrear:2020pzk}.

\section{Conclusions}
\label{sec:org087589e}
We have reported on the first neutron background measurement campaign carried out with newly developed, 3D nuclear recoil imaging gas time projection chambers, the BEAST TPCs, at SuperKEKB. We reported on both calibration of the energy scale and directional track reconstruction of nuclear recoil events. We provided measurements of the rate, energy spectra, and directional composition of nuclear recoils induced by fast neutron backgrounds during the first phase of SuperKEKB commissioning. We demonstrated two novel measurements and analysis techniques: First, 3D vector tracking of helium recoils was used to separate nuclear recoil events with positive radial velocity from those traveling in the opposite direction. Second, angular distributions of recoils were used to separate beam-gas and Touschek backgrounds. 

We have compared those measurements to predictions from dedicated beam background simulations and are able to make such comparisons for recoil energies down to 50 keV$_{\rm ee}$. This enabled first feedback on the Belle II / SuperKEKB neutron background simulation. Overall, we found reasonably good agreement between energy spectra and angular distributions of neutron recoils in simulation and measurements, suggesting that the beam optics simulation, neutron generation and accelerator material distribution are accurately simulated. However, several single beam background components appear incorrectly normalized. This discrepancy was also observed in other (non-neutron) early background studies at SuperKEKB~\cite{lewis19_first_measur_beam_backg_at_super}, and has since been improved via an improved simulation of beam collimators~\cite{andrii}.

We conclude that the performance of this system of TPCs holds significant promise for further background study in later SuperKEKB commissioning. More generally, TPC readout via pixel chips appears suitable also for future low-background experiments, especially low mass directional dark matter searches.

\section{Acknowledgements}
We thank the SuperKEKB group for accelerator operations and providing measurements of beam parameters. We thank John Kadyk, whose work on TPCs with pixel chip readout~\cite{Kim:2008zzi} inspired this project. We thank Maurice Garcia-Sciveres at Lawrence Berkeley National Laboratory for providing custom metalized pixel chips. We thank Marc Rosen and Kamaluoawaiku Beamer of the University of Hawaii for their assistance in designing and installing the BEAST II mechanical support structure. We thank Tommy Lam and Annam Lê for their help in constructing and testing the TPCs. We thank David-Leon Pohl and Jens Janssen from the University of Bonn for contributing the data acquisition software for the ATLAS FE-I4B pixel chips.  We acknowledge support from the U.S. Department of Energy (DOE) via Award Numbers DE-SC0007852, DE-SC0010504, via the U.S. Belle II Project administered by Pacific Northwest National Laboratory (DE-AC05-76RL01830), and via U.S. Belle II Operations administered by Pacific Northwest National Laboratory and Brookhaven National Laboratory (DE-SC0012704).

\bibliographystyle{elsarticle-num2}
\bibliography{phase1-tpcs}
\end{document}